\documentstyle[prd,eqsecnum,aps,floats]{revtex}
\def\NOBSERVED{41}
\def\NBKG{24.8}
\def\DBKG{2.4}
\def\EXCESS{16.2}
\def\XSEC{7.1}
\def\DXSEC{2.8}
\def\DXSECSYS{1.5}
\def\TLUM{110.3}
\def\psig{0.0006}
\def\sign{3.2}
\def\DOMASS{172.1}

\def\NNCUT{0.85}
\def\D0XSEC{5.6}
\def\DD0XSEC{1.8}
\def\Aplan{$\cal{A}$}
\def\Spher{$\cal{S}$}
\def\C{$\cal{C}$}
\def\met{\mbox{${\hbox{$E$\kern-0.6em\lower-.1ex\hbox{/}}}_T$}} 
\def\etarms{$\left<\eta^2\right>$} 
\def\ttbar{$t\bar{t}$}
\def\D0{D\O}
\def\etal{{\sl et al.}}                 
\input psfig
\begin{document}
\title {Measurement of the top quark pair production cross section in \\ 
$p\bar{p}$ collisions using multijet final states}
\author{                                                                      
B.~Abbott,$^{40}$                                                             
M.~Abolins,$^{37}$                                                            
V.~Abramov,$^{15}$                                                            
B.S.~Acharya,$^{8}$                                                           
I.~Adam,$^{39}$                                                               
D.L.~Adams,$^{48}$                                                            
M.~Adams,$^{24}$                                                              
S.~Ahn,$^{23}$                                                                
H.~Aihara,$^{17}$                                                             
G.A.~Alves,$^{2}$
% ...
\\                                                                            
\vskip 0.70cm                                                                 
\centerline{(D\O\ Collaboration)}                                             
\vskip 0.70cm                                                                 
}                                                                             
\address{                                                                     
\centerline{$^{1}$Universidad de Buenos Aires, Buenos Aires, Argentina}       
\centerline{$^{2}$LAFEX, Centro Brasileiro de Pesquisas F{\'\i}sicas,         
                  Rio de Janeiro, Brazil}                                     
\centerline{$^{3}$Universidade do Estado do Rio de Janeiro,                   
                  Rio de Janeiro, Brazil}                                     
\centerline{$^{4}$Universidad de los Andes, Bogot\'{a}, Colombia}             
\centerline{$^{5}$DAPNIA/Service de Physique des Particules, CEA, Saclay,     
                  France}                                                     
\centerline{$^{6}$Panjab University, Chandigarh, India}                       
\centerline{$^{7}$Delhi University, Delhi, India}                             
\centerline{$^{8}$Tata Institute of Fundamental Research, Mumbai, India}      
\centerline{$^{9}$Kyungsung University, Pusan, Korea}                         
\centerline{$^{10}$Seoul National University, Seoul, Korea}                   
\centerline{$^{11}$CINVESTAV, Mexico City, Mexico}                            
\centerline{$^{12}$Institute of Nuclear Physics, Krak\'ow, Poland}            
\centerline{$^{13}$Institute for Theoretical and Experimental Physics,        
                   Moscow, Russia}                                            
\centerline{$^{14}$Moscow State University, Moscow, Russia}                   
\centerline{$^{15}$Institute for High Energy Physics, Protvino, Russia}       
\centerline{$^{16}$University of Arizona, Tucson, Arizona 85721}              
\centerline{$^{17}$Lawrence Berkeley National Laboratory and University of    
                   California, Berkeley, California 94720}                    
\centerline{$^{18}$University of California, Davis, California 95616}         
\centerline{$^{19}$University of California, Irvine, California 92697}        
\centerline{$^{20}$University of California, Riverside, California 92521}     
\centerline{$^{21}$Florida State University, Tallahassee, Florida 32306}      
\centerline{$^{22}$University of Hawaii, Honolulu, Hawaii 96822}              
\centerline{$^{23}$Fermi National Accelerator Laboratory, Batavia,            
                   Illinois 60510}                                            
\centerline{$^{24}$University of Illinois at Chicago, Chicago,                
                   Illinois 60607}                                            
\centerline{$^{25}$Northern Illinois University, DeKalb, Illinois 60115}      
\centerline{$^{26}$Northwestern University, Evanston, Illinois 60208}         
\centerline{$^{27}$Indiana University, Bloomington, Indiana 47405}            
\centerline{$^{28}$University of Notre Dame, Notre Dame, Indiana 46556}       
\centerline{$^{29}$Purdue University, West Lafayette, Indiana 47907}          
\centerline{$^{30}$Iowa State University, Ames, Iowa 50011}                   
\centerline{$^{31}$University of Kansas, Lawrence, Kansas 66045}              
\centerline{$^{32}$Louisiana Tech University, Ruston, Louisiana 71272}        
\centerline{$^{33}$University of Maryland, College Park, Maryland 20742}      
\centerline{$^{34}$Boston University, Boston, Massachusetts 02215}            
\centerline{$^{35}$Northeastern University, Boston, Massachusetts 02115}      
\centerline{$^{36}$University of Michigan, Ann Arbor, Michigan 48109}         
\centerline{$^{37}$Michigan State University, East Lansing, Michigan 48824}   
\centerline{$^{38}$University of Nebraska, Lincoln, Nebraska 68588}           
\centerline{$^{39}$Columbia University, New York, New York 10027}             
\centerline{$^{40}$New York University, New York, New York 10003}             
\centerline{$^{41}$University of Rochester, Rochester, New York 14627}        
\centerline{$^{42}$State University of New York, Stony Brook,                 
                   New York 11794}                                            
\centerline{$^{43}$Brookhaven National Laboratory, Upton, New York 11973}     
\centerline{$^{44}$University of Oklahoma, Norman, Oklahoma 73019}            
\centerline{$^{45}$Brown University, Providence, Rhode Island 02912}          
\centerline{$^{46}$University of Texas, Arlington, Texas 76019}               
\centerline{$^{47}$Texas A\&M University, College Station, Texas 77843}       
\centerline{$^{48}$Rice University, Houston, Texas 77005}                     
}                                                                             
\maketitle
%
\begin{abstract}
We have studied $t\bar{t}$ production using multijet final states
in $p\bar{p}$ collisions at a center-of-mass energy of 1.8~TeV,
with an integrated luminosity of \TLUM~pb$^{-1}$. Each of the top quarks
with these final states decays exclusively to a bottom quark and a $W$ boson,
with the $W$ bosons decaying into quark-antiquark pairs. The analysis
has been optimized using neural networks to achieve the smallest expected
fractional uncertainty on the $t\bar{t}$ production
cross section, and yields a cross section
of \XSEC~$\pm$ \DXSEC~(stat)~$\pm$ \DXSECSYS~(syst)~pb, assuming a
top quark mass of \DOMASS~GeV/$c^2$.
Combining this result with previous D\O~measurements, where one
or both of the $W$ bosons decay leptonically, gives
a \ttbar~production cross section of 5.9~$\pm$ 1.2~(stat) $\pm$ 1.1 (syst)~pb.
\end{abstract}
%
\twocolumn
\tableofcontents
%
\newpage
\section {Introduction}

In the standard model, the top quark decays to a $b$ quark and a $W$ boson,
and the dominant decay of the $W$ boson is into a quark-antiquark pair.
Events with a $t\bar{t}$ pair can have both $W$ bosons
decaying to quarks. This is referred to
as the ``all-jets'' channel, and is expected to
account for 44\% of the $t\bar{t}$ production cross section.

The observation of top quark production \cite{d0obs,cdfobs} in the channels
involving one or two leptons motivates us to investigate $t\bar{t}$ decays into
other channels. D\O~has measured a top quark mass, $m_t$, of 
172.1 $\pm$ 5.2 (stat) $\pm$ 4.9 (syst) GeV/$c^2$~\cite{snydermass}
and a \ttbar~production cross section of 5.6~$\pm$ 1.4~(stat) $\pm$ 1.2~(syst) 
pb~\cite{meenajim}, while CDF has measured a mass of
175.9~$\pm$ 4.8~(stat) $\pm$ 4.9~(syst) GeV/$c^2$~\cite{cdfmass}
and a \ttbar~production cross section of 7.6 $^{+1.8}_{-1.5}$ pb~\cite{cdfxs}.
Recently, CDF has reported on the all-jets channel~\cite{cdfalljets}, and
finds the $t\bar{t}$ production cross section to be 10.1 $^{+4.5}_{-3.6}$ pb
and a top quark mass of 186 $\pm$ 10 (stat) $\pm$ 12 (syst) GeV/$c^2$.

The work presented here is based on 110.3~$\pm$ 5.8 pb$^{-1}$ of 
data recorded between August 1992 and February 1996 at the Fermilab Tevatron 
collider, with a $p\bar{p}$ center-of-mass energy of 1.8 TeV.
Assuming the branching ratio and cross section predicted by the standard
model, we expect approximately
200 $t\bar{t}\rightarrow$ all-jets 
events in this data sample.

The signature for $t\bar{t}$ production in the all-jets channel is
six or more high transverse momentum jets with kinematic properties consistent
with the top quark decay hypothesis.  At least two of these jets originate from
$b$ quarks. The background to this signature
consists of events from other processes that can also produce six
or more jets.  The $t\bar{t}$ channel is one of the few examples of multijet
final states that are dominated by quarks rather than gluons.
This fact has motivated us to include the characteristic differences
between quark and gluon jets in separating the top quark to all-jets signal
from background.

Interest in the all-jets decay channel of top quarks also stems from the
fact that, without any unobserved particles in the final-state, the
all-jets mode is the most kinematically constrained of all the top quark decay
channels.  Furthermore, since the top quark is quite massive, decays via charged
Higgs may be possible.  If channels such as $t \rightarrow H^{+} b$ have
a significant branching fraction, the main effect could be a deficit in
the $t\bar{t}$ final states with energetic electrons or muons, relative
to the all-jets channel.
%
%
\section{Outline of Method}
The search for the top quark in the all-jets channel began with the imposition 
of preliminary selection criteria at the trigger stage, followed by 
more stringent criteria in the offline analysis. As these initial
criteria were not very restrictive, the observed cross section,
primarily from QCD processes, was more than three thousand times larger than 
the expected signal. The principal
challenge in the search was to develop a set of selection criteria that could
significantly improve the signal-to-background ratio, and provide an
estimate of the background remaining after imposing any selection requirements. 

The data sample consisted of over 600,000 events after the initial selection 
criteria. Because of the small number of $t\bar{t}$ events expected in
the presence of this large
background, and with only modest discrimination in any single
kinematic or topological property, traditional methods of analysis
were inadequate.  The analysis would have to involve many variables,
which are likely to be highly correlated.
Neural networks were chosen as the
appropriate tool for handling many variables simultaneously.
	
The analysis relied on Monte Carlo simulations to model the properties
of $t\bar{t}$ events.  These simulations were performed for
different top quark masses, and the final results interpolated to the mass 
measured by the D\O~collaboration.
We note that the $t\bar{t}$ detection efficiency is not
strongly dependent on the assumed mass of the top quark.

In contrast, the background model was determined entirely
from data.  An advantage of the overwhelming background-to-signal ratio is that
the data provide an almost pure background sample. 
This approach obviates a number of concerns when calculating the
background.  The background is predominantly QCD multijet production, which
involves higher-order processes that may not be well modeled in a
Monte Carlo simulation.  Furthermore, detector effects are
implicitly included when data are employed for the model of the background.

Soft-lepton tagging, using muons embedded in jets, serves as a possible
signature for the presence of a $b$ quark within the jet, and
is referred to as $b$-tagging.  By identifying the muon from the 
semileptonic decay of a $b$ quark (or the sequential decay),
$b$-tagging of jets improves the signal-to-background ratio significantly.
The $t\bar{t}$ events are tagged roughly 20\% of the time,
whereas the tag rate for QCD multijet events with similar requirements
is about 3\%.  Requiring the presence of a muon tag in the
event therefore provides nearly a factor of ten in background rejection and
a method to estimate this background.

The background calculation relied on being able to predict the number of
events that are $b$-tagged, based on events without such tags.
To make the untagged data represent the background in this analysis, 
a way of estimating the tagging rate in QCD events was needed.  This was done
by constructing a ``tag rate'' function, determined from data, that is applied
to each jet separately. This function is simply the probability
for any individual jet to have a muon tag.
Application of the tag rate function to each jet in untagged events gives
the background model for our final event sample.
The presence of $t\bar{t}$ signal was identified by
an excess observed in the data above this background.  This excess should be 
small in the regions of the neural network output where background dominates,
but should be enhanced where significant signal is expected.

This analysis employed two neural networks to extract the final $t\bar{t}$ 
cross section. The first had as its input variables those parameters involving 
kinematic and topological properties of the events that were
highly correlated. The output of this neural network was used as an input
variable to a second neural network, along with three other inputs.
These three inputs were the transverse momentum ($p_T$)
of the tagging muon, a discriminant based on the widths of the jets, 
and a likelihood variable that parameterized the degree to which an event
was consistent with the $t\bar{t}$ decay hypothesis.  These three variables 
were less correlated than the kinematic variables used in the first neural 
network. The $t\bar{t}$ cross section was determined from the output of this 
second neural network by fitting the neural network output distributions of 
the signal and background outputs to the observed data.
%
\section{The D\O~Detector}

\begin{figure}
\centerline{\psfig{figure=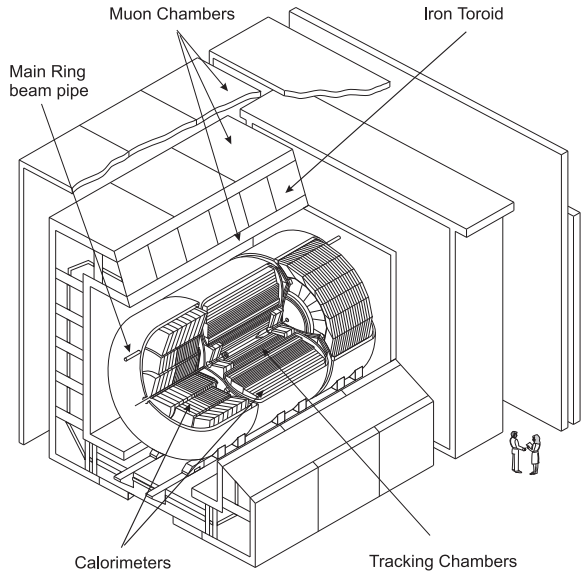,width=3in}}
\caption{\small Isometric view of the D\O~detector.}
\label{cutaway}
\end{figure}

D\O~is a multipurpose detector designed to study $p\bar{p}$ collisions
at the Fermilab Tevatron Collider. The detector 
was commissioned during the summer of 1992.  A full description
of the detector can be found in references \cite{d0detector,ouroldprd}. 
Here we describe the properties of the detector that are most
relevant to the search in the all-jets channel. An isometric
view of the detector is shown in Fig.~\ref{cutaway}.
 

%
\subsection{Tracking system}

The tracking system consists of a vertex drift chamber,
a transition radiation detector, a central drift chamber, and
two forward drift chambers. The system provides
charged-particle tracking over the pseudorapidity region $|\eta|<$ 3.2,
where $\eta$ = $-\ln({\tan{(\theta/2)}})$; $\theta$ and $\phi$ are, 
respectively, the polar and azimuthal angles relative to the proton beam axis. 
The resolution for charged particles is 2.5 mrad in $\phi$ and 28 mrad in 
$\theta$. The position of the interaction vertex along the beam direction 
($z$) is determined typically to an accuracy of 8 mm.
%
\subsection{Calorimeter}
   
The liquid-argon calorimeter, using uranium and stainless steel/copper absorber,
is divided into three parts: a central calorimeter
and two end calorimeters. Each part consists of an inner electromagnetic
section, a fine hadronic section, and a coarse hadronic section, housed in a
stainless steel cryostat. The intercryostat detector
consists of scintillator tiles
inserted in the space between the central and end calorimeter 
cryostats. 
In addition, ``massless gaps'',
installed inside both central and end calorimeters, are
active readout cells, without absorber material, located inside the
cryostat adjacent to the cryostat walls.
The intercryostat detector and massless gaps
improve the energy resolution for jets that straddle two cryostats. The
calorimeter covers the pseudorapidity range $|\eta|<$ 4.2, and has a typical 
segmentation of
0.1 $\times$ 0.1 in $\Delta\eta$ $\times$ $\Delta\phi$.  The energy
resolution is $\delta(E)/E$ = 15\%/$\sqrt{E{\rm(GeV)}}$ $\oplus$ 0.4\% for 
electrons. For charged pions, the resolution is approximately 
50\%/$\sqrt{E{\rm(GeV)}}$, and for jets approximately 
80\%/$\sqrt{E{\rm(GeV)}}$ \cite{d0detector,ouroldprd}. 

As can be seen in Fig.~\ref{cutaway}, the Main Ring beam pipe penetrates
the outer hadronic section of the calorimeters and the muon spectrometer.
The Main Ring carries protons with energies between 8-150 GeV, and is
used in antiproton production during the Tevatron $p\bar{p}$ running.
Because of this, any losses from the Main Ring can produce
backgrounds in the detector that must be removed.
%
\subsection{Muon spectrometer}
The D\O~experiment detects muons using proportional drift tubes (PDT) and an 
iron toroid.  Because muons from top quark decays populate predominantly the 
central region, this analysis uses muon detection systems in the region 
$|\eta|<$ 1. 

The combined material in the calorimeter and iron toroid has between
13 and 19 interaction lengths (the range-out energy for muons is approximately
3.5 GeV), making background from hadronic punchthrough
negligible. Also, the small central tracking volume minimizes background
from in-flight decays of pions and kaons.

A typical muon track is measured in four layers of PDTs before,
and six layers after, the iron toroid. The six layers are constructed 
in two super-layers that are separated by about one meter to provide
a good lever arm for measuring the muon momentum, $p$.
The muon momentum is determined from its deflection angle in the magnetic
field of the toroid. The momentum resolution is limited by multiple 
scattering in the traversed material, knowledge of the 
integrated magnetic field,
and resolution on the measurement of the deflection angle. The resolution
is roughly Gaussian in 1/$p$, and is approximately
$\delta(1/p)$ = 0.18($p$-2)/$p^2\oplus$0.003 (with $p$ in GeV/$c$) for
the algorithms that were used in this analysis.
%
\section{Data Sample}

This section describes the data sample and the simulated events for the
\ttbar~signal used in our analysis.

\subsection{Initial selection criteria}
\label{TRIGGERS}

The data sample was selected by imposing both hardware 
(Level 1) and software (Level 2) trigger requirements.
These requirements were modified slightly over the course of the 1992--1996 run
in order to accommodate the higher instantaneous luminosities later in the run.
Table I indicates the three main running periods, the run numbers
associated with these periods, and the integrated luminosity collected.

\begin{table}
 \begin{small}
 \caption{\small Main running periods of the 1992--1996 run.}
 \begin{center}
 \begin{tabular}{l|c|c|c}
 Run            &                 &  Run        & Integrated  \\
 Period         &  Dates          &  Numbers    & Luminosity  \\
 \hline
 Ia             & 1992--1993       & 50000--70000   & 13.0 pb$^{-1}$ \\
 Ib             & 1993--1995       & 70000--94000   & 86.4 pb$^{-1}$ \\
 Ic             & 1995--1996       & 94000--96000   & 10.8 pb$^{-1}$ \\
 \end{tabular}
 \end{center}
 \end{small}
\end{table}

The hardware trigger required the presence of at least four
calorimeter trigger towers 
(0.2 $\times$ 0.2 in $\Delta\eta$ $\times$ $\Delta\phi$),
each with transverse energy $E_T >$ 5 GeV, for the Ia period.
In the Ib and Ic periods, the $E_T$ requirement was raised to 7 GeV, and
an additional requirement for at least three large tiles 
(0.8 $\times$ 1.6 in $\Delta\eta$ $\times$ $\Delta\phi$)
with  $E_T >$ 15 GeV was imposed.  These were imposed to reduce the
trigger rate and avoid saturating the bandwidth of the trigger system at
high instantaneous luminosities ($\geq$ 10$^{31}$ cm$^{-2}$ s$^{-1}$).

The software filter required five jets, defined by
$\cal{R}$=$\sqrt{(\Delta\eta)^2+(\Delta\phi)^2}$=0.3 cones,
with $|\eta| <$ 2.5 and $E_T >$ 10 GeV. Again, in order to reduce the data 
rate at high luminosities during the Ib period, a further condition was added 
requiring the scalar sum of the $E_T$ of all jets (defined as $H_T$)
to be greater than 110 or 115 GeV, depending upon run number.
This $H_T$ requirement was raised to 120 GeV during the Ic period.
The effects of these changes on the acceptance for $t\bar{t}$ events were 
studied using Monte Carlo simulations, and were found to be negligible.

\begin{figure}
\centerline{\psfig{figure=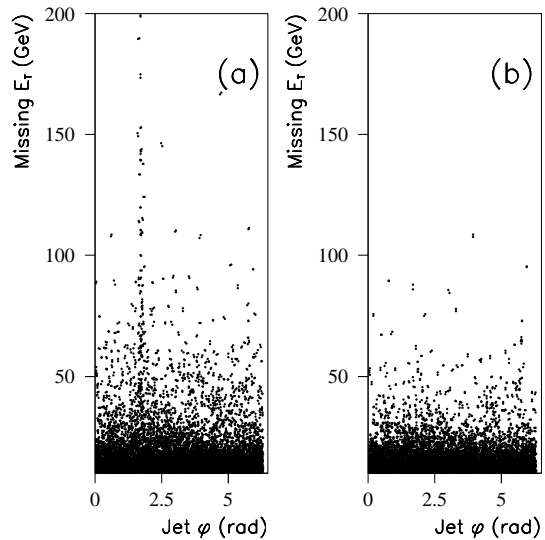,width=3in}}
\caption{\small The effect of imposing requirements to reject Main Ring events.
 A scatter plot of missing $E_T$ versus $\phi$ for jets before
(a), and after (b), imposing our Main Ring requirements.}
\label{mainring}
\end{figure}

In addition to imposing trigger and filter requirements, a set of offline
selection criteria was used to reduce the data sample to a manageable size
without greatly affecting the acceptance for the \ttbar~signal. First, $H_T$ was
required to be greater than 115 GeV, where the sum used ${\cal{R}}$=0.5 jets
with $|\eta|<$ 2.5 and $E_T >$ 8 GeV. Also, requirements were imposed
in order to eliminate events with spurious jets due to spray from the Main Ring
or effects from noisy cells in the calorimeter \cite{Cathythesis,Eunilthesis}. 
For example, Fig.~\ref{mainring} shows the imbalance in transverse energy, or 
missing $E_T$ ($\met$), in the event versus the azimuthal angle ($\phi$) of the
jet, before and after the rejection of Main Ring events. We see that our 
requirements have removed the spurious cluster of jets in the region where the 
Main Ring pierces the D\O~detector (1.6 $< \phi <$ 1.8).  Table II summarizes 
the impact of the trigger and initial reconstruction criteria on the 
\ttbar~signal for a top quark mass of 180 GeV/$c^2$.

\begin{table}
 \begin{small}
 \caption{\small Initial criteria used for data selection.}
 \begin{center}
  \begin{tabular}{l|c|c|c}
             &                &             & Cumulative \\
             &                &  Effective  & Efficiency  \\
 General     & Sequential     &  cross      & ($m_t$=180  \\
 Conditions  & Requirements   &  section    & GeV/$c^2$)  \\
 \hline
 Level 1          & Four trigger towers  &  &  \\
 trigger          & $E_T>5,7$ GeV (Ia, Ib-c) & 0.4 $\pm$ 0.1 $\mu$b &  0.98 \\
                  & Three large tiles &  &  \\
                  & $E_T>15$ GeV  (Ib-c) &  &  \\
 \hline
 Level 2          & Five ${\cal{R}}$=0.3 jets &  &   \\
 filter           & $|\eta|<$2.5, $E_T>10$ GeV & 20 $\pm$ 5 nb    & 0.92  \\
                  & $H_T>110,115$ GeV (Ib) & &  \\
                  & $H_T>120$ GeV (Ic) & &  \\
 \hline
                & $H_T>115$ GeV from & &   \\
 Offline        & ${\cal{R}}$=0.5 jet cones & &   \\
                & $|\eta|<$2.5, $E_T>8$ GeV & 5.4 $\pm$ 1.3 nb & 0.87  \\
                & Cuts for spurious jets &  &  \\
 \end{tabular}
 \end{center}
 \end{small}
\end{table}

%
\subsection{Jet algorithms}

The jet algorithm is the fundamental analysis tool in the search for
$t\bar{t}$ events in the all-jets mode.  One of the most important
considerations in choosing a jet algorithm is the efficiency
for reconstructing the six primary $t\bar{t}$ decay products.
The $\eta$ distribution of the jets from $t\bar{t}$ decays tends to be
quite narrow, and therefore the $\cal{R}$ separation between 
adjacent jets is frequently small.  When two jets are too close together, 
they may not be resolved, leading to reconstruction inefficiency. 

\begin{figure}
\centerline{\psfig{figure=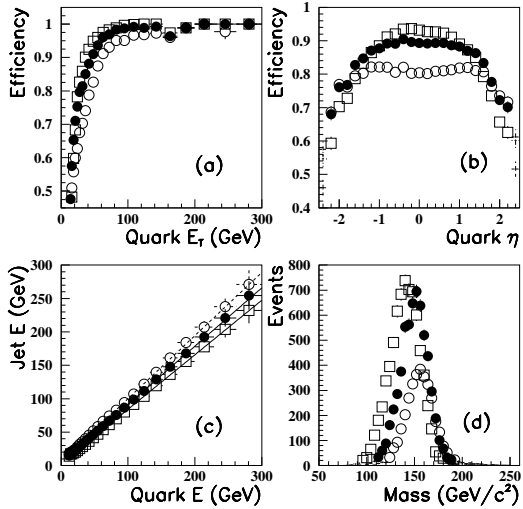,width=3in}}
\caption{
Jet reconstruction for $t\bar{t}$ Monte Carlo events
({\sc herwig}, $m_t$ = 175 GeV/$c^2$) for various cone sizes:
${\cal{R}}$=0.3 (open squares), ${\cal{R}}$=0.5 (full circles), and 
${\cal{R}}$=0.7 (open circles).
(a) Jet finding efficiency versus quark $E_T$.
(b) Jet finding efficiency versus quark $\eta$.
(c) Reconstructed jet energy versus that of the input quark.
(d) Reconstructed mass of the top quark from correct jet combinations,
where the areas reflect the relative efficiencies.}
\label{algcomp}
\end{figure}

Figure~\ref{algcomp} shows the reconstruction efficiency for the cone
jet algorithm~\cite{conealg} with various cone sizes
for simulated $t\bar{t}$ events in the all-jets
channel, as generated with the {\sc herwig} Monte Carlo program\cite{herwig}.
Here, the definition of a quark includes any final state gluon radiation 
added back to the quark momentum. The matching of reconstructed jets to 
quarks relies on using combinations of the two that minimize the distance in
$\cal{R}$ between them. A jet is considered to be matched only if
that distance is less than $\Delta\cal{R}$=0.5, the energy of the jet is
within a factor of two of the quark energy, and the
reconstructed jet $E_T$ is greater than 10 GeV.

Figures~\ref{algcomp}(a) and \ref{algcomp}(b) show how the reconstruction
efficiency depends on quark $E_T$ and
$\eta$ for the cone algorithm with different cone sizes.
The $\cal{R}$=0.3 cone algorithm shows a higher 
jet reconstruction efficiency than the larger cone algorithms. 
In the central region, the $\cal{R}$=0.3 cone algorithm has an efficiency
of 94\%, while the $\cal{R}$=0.5 and $\cal{R}$=0.7
cone algorithms are 90\% and 81\% efficient, respectively.
Given an average efficiency $\epsilon$ for reconstructing a single jet, the
reconstruction efficiency for finding $t\bar{t}$ events (with six or more jets)
will be of the order of $\epsilon^6$. Therefore, larger cone sizes are
less efficient in the multijet environment.

Figure~\ref{algcomp}(c) shows the correspondence between parton
and jet energies found for various cone algorithms, after D\O~jet energy
corrections are applied (see next section). Linear fits to the quark-jet 
correlation in energy are shown in Fig.~\ref{algcomp}(c) for the three cone 
algorithms. Figure~\ref{algcomp}(d) shows the three-jet invariant mass for the 
correct combinations of jets matching top and antitop quarks. The areas of the 
mass distributions reflect the event reconstruction efficiencies for different 
algorithms.

The shift in the reconstructed mass from the input mass of the top quark
(175 GeV/$c^2$) shows that the jet algorithms are not equivalent. The shift 
in three-jet mass from the nominal input top quark mass increases as the 
cone radius is decreased. The widths of the mass distributions are not very 
sensitive to the choice of cone size.  The overall root-mean-square, {\sc rms},
spread in reconstructed mass for correct combinations of jets is approximately 
10\% of the mass.

In summary, there are two competing effects when choosing the optimal jet cone 
size.  Smaller cone sizes are better able to resolve separate jets, but do not 
do as well at reconstructing jet energy. However, the ability to resolve 
individual jets was deemed of higher importance in the search for a signal.
Hence the $\cal{R}$=0.3 cone algorithm is preferred for analyzing multijet 
events. But, due to the relatively large shift in the jet energy
for the $\cal{R}$=0.3 cone algorithm, we chose to use the $\cal{R}$=0.5 cone 
algorithm for calculating some quantities that emphasize energy response at the 
expense of jet efficiency. Jets with  $E_T~<$ 8 GeV, before application of 
energy corrections (see Sec.~IV.C), were discarded.
%
\subsection{Jet energy correction}
\label{jetenergycorr}
D\O~has developed a correction procedure~\cite{cafix} to calibrate jet energies,
which is applied to both data and Monte Carlo. The underlying assumption is 
that the true jet energy, $E_{\rm ptcl}$, is the sum of the energies of all
final state particles entering the cone algorithm applied at the calorimeter 
level. $E_{\rm ptcl}$ is obtained from the energy measured in the calorimeter, 
$E_{\rm meas}$, as follows:
\begin{eqnarray}
  E_{\rm ptcl} = 
  \frac{E_{\rm meas}- 
  E_{\rm O}({\cal R},\eta,{\cal L})}
  {R(\eta,E,{\sc rms})
  S({\cal R},\eta,E)} \, , 
\end{eqnarray}
\noindent
where:

\begin{itemize}

\item $E_{\rm O}({\cal R},\eta,{\cal L})$
is an offset, which includes the physics of the underlying event, noise from
the radioactive decay of the uranium absorber,
the effect of previous crossings (pile-up), and the 
contribution of additional contemporaneous $p\bar{p}$ interactions.
The physics of the underlying
event is defined as the energy contributed by spectators to the hard parton
interaction which resulted in
the high-$p_{T}$ event. 
This offset increases as a function of the cone size ${\cal R}$.
It also depends on $\eta$ and on the instantaneous luminosity, ${\cal L}$,
which is related to the contribution from
the additional $p\bar{p}$ interactions. 

\item $R(\eta,E,{\sc rms})$ is the energy response of the calorimeter.  
It is nearly independent of the jet cone size, ${\cal R}$, but does depend 
on the {\sc rms} width of the jet. The width dependence accounts for 
differences in the calorimeter response to narrow jets, which fragmented 
into fewer particles (of, on average, higher energy) than broader jets, 
with larger particle multiplicities. Because the various detector components 
are not identical, $R$ also depends on detector $\eta$. $R$ is typically less 
than one, due to energy loss in the uninstrumented regions between modules, 
differences between the electromagnetic ($e$) and hadronic response ($h$) of 
the detector ($e/h>1$), and module-to-module inhomogeneities. 

\item {\bf $S({\cal R},\eta,E)$} is the fraction of the jet
energy that is deposited inside the algorithm cone. Since the jet energy is
corrected back to the particle level, the effects of calorimeter showering 
must be removed. $S$ is less than one, meaning that the effect of
showering is a net flux of energy from inside to outside the cone.
$S$ depends strongly on the cone size ${\cal R}$, energy, and $\eta$. 

\end{itemize}

%
\subsection{Characteristics of jets}
\begin{figure}
\centerline{\psfig{figure=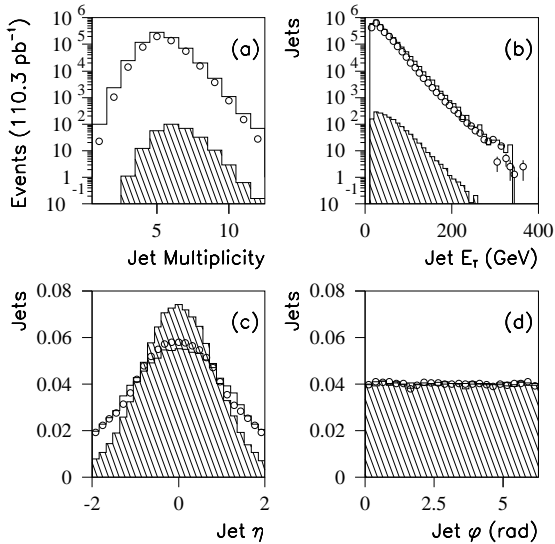,width=3in}}
\caption{ Properties of jets with $\cal{R}$=0.3 cones. Data from
the~Ia (histograms) and Ib (circles) periods, and \ttbar~{\sc herwig} for
$m_t$=175 GeV/$c^2$ (shaded histograms). Only jets with  $E_T$ $>$ 10 GeV and 
$|\eta|<2$ are included. Distributions in (a) jet multiplicity and (b) jet 
$E_T$ are each normalized to the expected number of events in 110.3 pb$^{-1}$ 
of data, while distributions in (c) jet $|\eta|$ and (d) jet $\phi$ are 
normalized to the same area.}
\label{jetprop}
\end{figure}

Comparisons of jet properties (jet multiplicity, inclusive jet $E_T$,
$\eta$, and $\phi$, for $\cal{R}$=0.3 cones)
are shown in Fig.~\ref{jetprop} for data from the Ia and Ib periods 
(see Table I) and for
$t\bar{t}$ Monte Carlo.
Only jets with $E_T >$ 10 GeV and $|\eta|<2$ are
included in the comparison. The results from Ia and Ib are in good agreement,
although Ib typically had higher instantaneous luminosity.

Figure~\ref{jetprop}(a) shows that for events with six jets,
the background ({\rm i.e.}, data) is at least three orders of magnitude
larger than the expected $t\bar{t}$ signal. The peak at five jets is the 
result of the initial event selection (see Table II). The inclusive jet $E_T$ 
spectrum in Fig.~\ref{jetprop}(b) falls exponentially at about the same rate 
for signal as for data, and the signal is consistently three orders of
magnitude below the data. In Fig.~\ref{jetprop}(c), the distributions of
jet $\eta$ are normalized to the same area for signal and data. The signal 
is concentrated in the central region, while the data extend to higher $\eta$.
There is a difference of the order of 10\% between~Ia and Ib in the 
intercryostat region ($|\eta| \approx$ 1.2) due to improvements in the Ib 
period. Figure~\ref{jetprop}(d) shows that the $\phi$ distribution of jets is
isotropic, except for a 5\% suppression in the region of the Main Ring.
The Monte Carlo does not simulate the effects of the
Main Ring, and consequently has no apparent structure in $\phi$.

\subsection{Simulation of $t\bar{t}$ events}
\label{topsim}
The simulation of $t\bar{t}$ events plays an important role in
extracting a signal in the presence of significant
background. It is necessary, therefore, to have a
good description of the production and decay of $t\bar{t}$ events
in order to calculate detector acceptances accurately and to develop
methods to identify  $t\bar{t}$ events in the data.

The $t\bar{t}$ events were generated for top quark masses between 120 and
220 GeV/$c^2$ for the reaction $p\bar{p} \rightarrow t\bar{t} + X$
using {\sc herwig}  as a primary model and {\sc isajet} \cite{isajet} as a 
check. The underlying assumptions in the fragmentation of partons are
different in the two programs. The generated events were put through the 
D\O~shower library~\cite{RAJA}, a fast detector simulation package
based on {\sc geant} \cite{GEANT}, which contains the effects of 
cracks and other dead material in the D\O~calorimeter, and provides 
accurate shower simulation. The {\sc geant} simulation has been tuned 
to achieve a good match between generated single-particle characteristics 
and observed data \cite{NWA}. Events were subsequently digitized, passed 
through the D\O~reconstruction program \cite{d0detector}, and subjected 
to the same selection criteria as the data (see Table II). Events passing 
these criteria served as the model for our studies of $t\bar{t}$ properties.

\begin{figure}
\centerline{\psfig{figure=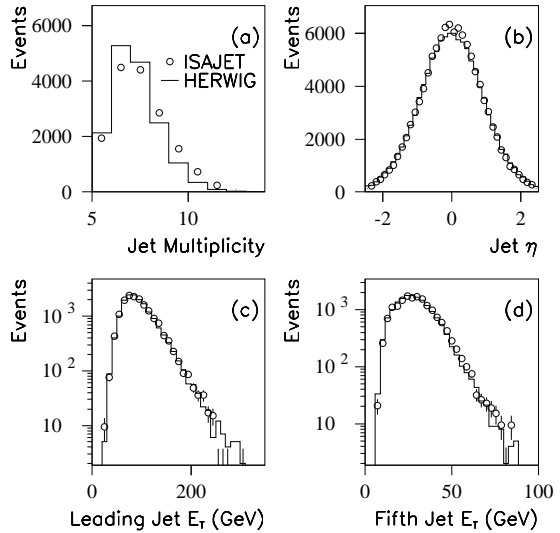,width=3in}}
\caption{ Comparisons of {\sc isajet} (circles) and
{\sc herwig} (histograms) for an
input top quark mass of 175 GeV/$c^2$, and jets with $\cal{R}$=0.3 cones, 
for (a) jet multiplicity,   (b) jet $\eta$, (c) $E_T$ of leading jet, and
(d) fifth highest jet $E_T$.  Bars on the points indicate statistical 
uncertainties (similar uncertainties, although not shown, apply for the 
histograms). The results from {\sc isajet} and {\sc herwig} in (a)--(d) 
are normalized to the same area.}
\label{dattbar}
\end{figure}

Generally, acceptances for $t\bar{t}$ production as calculated with 
{\sc herwig} or {\sc isajet} agree to within 10\%, and any differences 
between the two are incorporated in the final systematic uncertainties.
As an illustration of the discrepancies, we show in Fig.~\ref{dattbar} 
distributions of jet multiplicity, jet $\eta$, the $E_T$ of the leading jet, 
and the fifth highest jet $E_T$ for {\sc herwig} and {\sc isajet}.  Except 
for jet multiplicity, these distributions are in good agreement. It has been 
shown \cite{ouroldprd} that {\sc isajet} produces more gluon radiation than
{\sc herwig}, in accord with our results in Fig.~\ref{dattbar}(a).

%
\section{Kinematic Parameters}
\label{kinematicvariables}

The principal background to the $t\bar{t}$ signal is QCD multijet production, 
which is dominated by a $2 \rightarrow 2$ parton process with additional
jets produced through gluon radiation.  Therefore, the background tends to have
jets that are more forward-backward in rapidity. The additional jets are 
generally lower in $E_T$ ({\rm i.e.}, softer) than the initial outgoing parent 
partons.  Furthermore, this extra radiation tends to lie in
a plane formed by the incoming beam and the two leading jets.

Because the mass of the top quark is large, the characteristic energy scale
(commonly called $Q^2$) of the $t\bar{t}$ event is significantly larger than 
that of the average QCD background event.  This means that $t\bar{t}$ events 
generally have jets with higher $E_T$, and have larger multijet invariant 
masses.

Extracting a signal from data dominated by background requires
the use of global kinematic parameters based on these differences.
Employing such parameters helps to differentiate between
the $t\bar{t}$ signal and background. We can summarize the salient 
features of the background, relative to the $t\bar{t}$ signal, as follows:

\begin{itemize}
\item The overall energy scale is lower; leading jets have lower
$E_T$; multijet invariant masses are smaller.
\item The additional radiated jets are softer (have lower $E_T$).
\item The event shape is more planar (less spherical).
\item The jets are more forward-backward in rapidity (less central).
\end{itemize}

We defined
two or more kinematic parameters that quantified aspects of each property.
Only the most effective of these were used and these are discussed below.
We found that, in general, better discrimination was achieved using
$\cal{R}$=0.3 cone jets (with $|\eta|<2.0$ and $E_T > 15$ GeV) than 
$\cal{R}$=0.5 cone jets. However, in some instances, $\cal{R}$=0.5 cone 
jets were used, and this is noted where it occurs.

Although correlations exist between many of the kinematic parameters, each 
includes useful information not fully contained in any of the others. These 
correlations are presented in Sec.~VI.D.
%
\subsection{Parameters sensitive to energy scale}

Any parameter that depends on the energy scale of the jets
is also sensitive to the mass of the top quark. These ``mass sensitive'' 
parameters usually provide better discrimination against QCD background 
than other parameters that provide only a measure of some topological 
feature. Three mass sensitive parameters are:

\begin{enumerate}
\setcounter{enumi}{0}

\item $H_T$ \\
The sum of the transverse energies of jets in a given event characterizes
the transverse energy flow, and is defined as:
\begin{equation}
H_T = \sum_{j=1}^{N_{\rm jets}} E_{T_j}
\end{equation}
where $E_{T_j}$ is the transverse energy of the $j$th jet, as ordered in
decreasing jet $E_T$ rank, and $N_{\rm jets}$ is the number of jets 
in the event.

\item $\sqrt{\hat{s}}$ \\
This parameter is the invariant mass of the $ N_{\rm jets}$ system.

\item $E_{T_1}$/$H_T$ \\
$E_{T_1}$ is the transverse energy of the $\cal{R}$=0.5 cone jet with highest 
$E_T$. This parameter characterizes the $E_T$ fraction carried by the leading 
jet, and tends to be high for QCD background. The $t\bar{t}$ events are likely 
to have transverse energy roughly equipartitioned among all six jets, and 
hence the leading $E_T$ jet is, on average, fractionally softer.

\end{enumerate}

\begin{figure}
\centerline{\psfig{figure=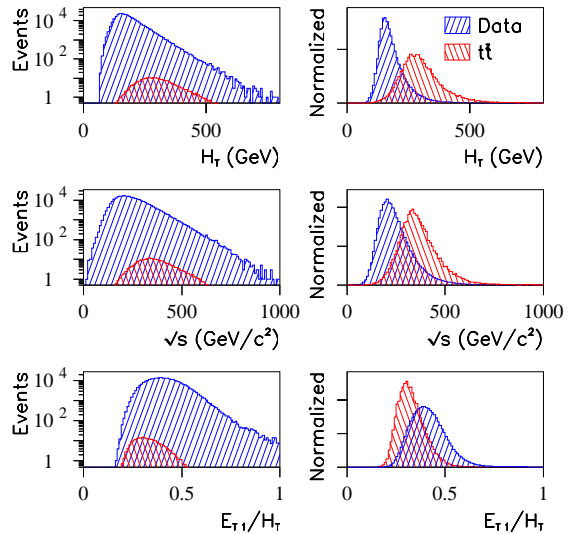,width=3in}}
\caption[The escale distributions.]
{\small The $H_T$, $\sqrt{\hat{s}}$, and $E_{T_1}$/$H_T$ distributions for 
data (predominantly background) and for {\sc herwig} $t\bar{t}$ generated at 
a top quark mass of 175 GeV/$c^2$. Each plot on the left is normalized 
according to the expected number of events. On the right the plots are 
normalized to unity and reveal significant discrimination between signal 
and background.}
\label{escale}
\end{figure}

Figure~\ref{escale} shows the distributions of $H_T$, $\sqrt{\hat{s}}$, and 
$E_{T_{1}}$/$H_T$, each of which reveals significant discrimination between 
signal and background. This and subsequent figures for the parameters are 
shown both normalized to cross section, and normalized to unity.

%
\subsection{Parameters sensitive to additional radiation}

As previously noted, the QCD background is primarily a $2 \rightarrow 2$ 
parton process that contains additional radiated gluons.  These gluons tend 
to be much softer than the leading partons, and therefore the jets associated 
with this radiation tend to have smaller $E_T$.  Three parameters that measure
the hardness of this radiation are:

\begin{enumerate}
\setcounter{enumi}{3}

\item $H_T^{3j}$ \\
This variable is defined as\cite{Cathythesis,Eunilthesis}:
\begin{equation}
H_T^{3j} = H_T - E_{T_1} - E_{T_2}
\end{equation}
where $E_{T_1}$ and $E_{T_2}$ are the transverse energies of the
two leading (highest $E_T$) jets.  By subtracting the $E_T$ of
the two leading jets, what remains is a better measure of any additional
gluon radiation in QCD events, enhancing the discrimination
between $t\bar{t}$ signal and QCD background.

\item $N_{\rm jets}^A$ \\
An average jet count parameter, $N_{\rm jets}^A$, provides a way to 
parameterize the number of jets in an event, while taking account of the 
hardness of these jets. We define:

\begin{equation}
N_{\rm jets}^{A} = { {\displaystyle \int^{55}_{15} }
                    E_{T}^{\rm thr} N(E_{T}^{\rm thr}) \, dE_{T}^{\rm thr}
           \over
                {\displaystyle \int^{55}_{15} }  E_{T}^{\rm thr} \, 
                 dE_{T}^{\rm thr} }
\end{equation}

where $N(E_{T}^{\rm thr})$ is the number of jets in a given event with 
$|\eta|<2.0$ and $E_T$ greater than some threshold, $E_T^{\rm thr}$ in GeV.
Therefore, this parameter corresponds to the number of jets, but is
more sensitive to jets of higher $E_T$ than just a simple jet count above
some given threshold. 

\item $E_{T_{5,6}}$ \\
The transverse energies of the fifth jet, $E_{T_5}$, and sixth jet, $E_{T_6}$, 
are also useful in discriminating QCD background from $t\bar{t}$ events.  
Our final selection (see Sec.~VII(a)) requires at least six jets. For 
background these usually correspond to soft radiation. The variable chosen is:

\begin{equation}
E_{T_{5,6}} = \sqrt{E_{T_5}~E_{T_6}}.
\end{equation}

\end{enumerate}

\begin{figure}
\centerline{\psfig{figure=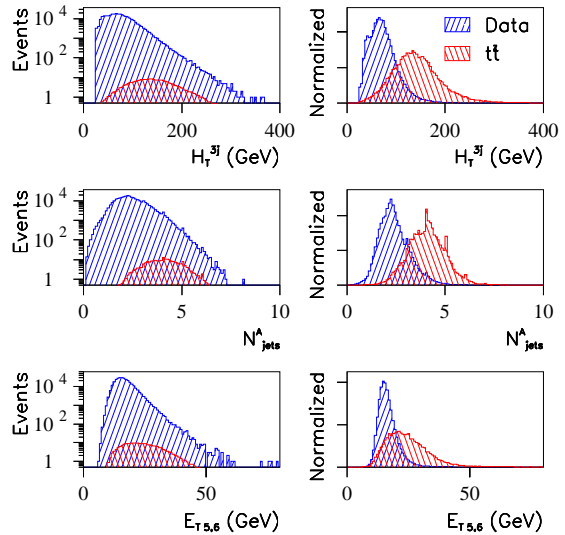,width=3in}}
\caption[Additional Rad. distributions]
{\small The $H_T^{3j}$, $N_{\rm jets}^A$, and $E_{T_{5,6}}$ distributions 
for data (predominantly background) and for {\sc herwig} $t\bar{t}$ events.  
Each distribution is normalized to the expected number of events (left) 
and to unity (right).}
\label{addrad}
\end{figure}

Figure~\ref{addrad} shows distributions of $H_T^{3j}$, $N_{\rm jets}^A$ and 
$E_{T_{5,6}}$. Again, these variables are effective in differentiating 
between signal and background.

%
\subsection{Aplanarity and sphericity}

The direction and shape of the momentum flow of jets in
$t\bar{t}$ production are different from those in
QCD background.  These differences can be quantified
using event-shape parameters\cite{smhb}. For each event, we define the
normalized momentum tensor $M_{ab}$:

\begin{equation}
 M_{ab} = \sum_{j}^{N_{\rm jets}} p_{ja}p_{jb}/\sum_{j}^{N_{\rm jets}} p_{j}^2
\end{equation}

\noindent
where $a$ and $b$ run over the $x,y,z$ components (indices of the tensor), and 
$j$ runs over the number of jets in an event. As is clear from its definition, 
$M_{ab}$ is a symmetric matrix that is always diagonalizable, and has
positive-definite eigenvalues ($Q_1,Q_2,Q_3$) satisfying the conditions:
\begin{equation}
Q_1 + Q_2 + Q_3 = 1~~~{\rm and}~~~0\leq Q_1\leq Q_2\leq Q_3.
\end{equation}
The equation $Q_1 + Q_2 + Q_3 = 1$ represents a plane in a space spanned by
$Q_1,Q_2$, and $Q_3$, and the inequality restricts
the range of each eigenvalue, as shown in Fig.~\ref{eigen}:

\begin{figure}
\centerline{\psfig{figure=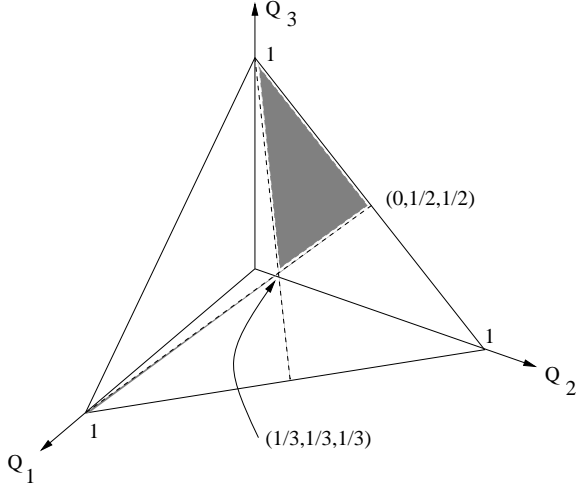,width=3in}}
\caption[The range of eigenvalues in the space spanned by $Q_i$s.]
{\small The allowed range of normalized momentum tensor eigenvalues in the 
space spanned by the $Q_i$.}
\label{eigen}
\end{figure}

\begin{eqnarray}
& & 0 \leq Q_1 \leq \frac{1}{3},   \\ \nonumber
& & 0 \leq Q_2 \leq \frac{1}{2},   \\ \nonumber
& & \frac{1}{3} \leq Q_3 \leq 1.   \\ \nonumber
\end{eqnarray}

The magnitude of any $Q_i$ represents the portion of momentum flow in the
direction of the $i^{th}$ eigenvector.
Limiting event shapes can therefore be characterized as follows:

\begin{itemize}
\item Linear : $Q_1 = Q_2 = 0$ and $Q_3$ = 1.
\item Planar : $Q_1 = 0$ and $ Q_2 = Q_3$ = $\frac{1}{2}$.
\item Spherical : $Q_1 = Q_2 = Q_3$ = $\frac{1}{3}$.
\end{itemize}

The aplanarity (\Aplan) and sphericity (\Spher) parameters that we use
are defined as follows:

\begin{enumerate}
\setcounter{enumi}{6}

\item ${\cal{A}} = \frac{3}{2} Q_1$,
\item ${\cal{S}} = \frac{3}{2} \left( Q_1+Q_2 \right)$,

\end{enumerate}
\noindent
with $0\leq {\cal{A}} \leq 0.5$ and $0\leq{\cal{S}} \leq 1$.

\begin{figure}
\centerline{\psfig{figure=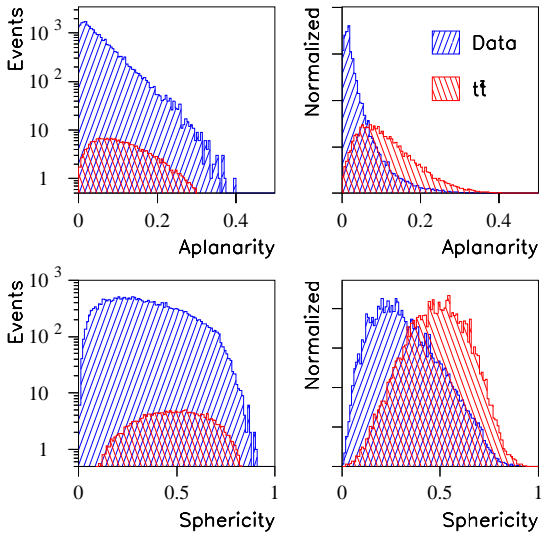,width=3in}}
\caption[The aplanarity and sphericity distributions.]
{\small The aplanarity and sphericity distributions for data
(predominantly background), and for {\sc herwig} $t\bar{t}$ events.    
Each distribution is normalized to the
expected number of events (left) and to unity (right).}
\label{aplsph}
\end{figure}

Top quark ($t\bar{t}$) events tend to have higher aplanarity and sphericity
than background events. We calculate \Aplan~and \Spher~in the $p\bar{p}$ 
collision frame; little difference is found using the parton center of mass 
frame. Distributions of \Aplan~and \Spher~for {\sc herwig} $t\bar{t}$
events for $m_t$=175 GeV/$c^2$ and for data are shown in Fig.~\ref{aplsph}.
%
\subsection{Parameters sensitive to rapidity distributions}

\begin{enumerate}
\setcounter{enumi}{8}

\item ${\cal{C}}$ \\
The centrality (${\cal{C}}$) parameter is defined as:
\begin{equation}
{\cal{C}} = \frac{H_T}{H_E},
\end{equation}
where
\begin{equation}
H_E = \sum_{j=1}^{N_{\rm jets}} E_{j}.
\end{equation}
Centrality is similar to $H_T$, characterizing the transverse energy in events,
but is normalized in such a way that it depends only weakly on the mass of the
top quark.

\item $\left<\eta^2\right>$ \\
To good approximation, the $\eta$ distribution for jets in $t\bar{t}$ events is
normally distributed about zero with an {\sc rms}, $\sigma_{\eta}$, close to 
unity.  With typically six or more jets in an event, the {\sc rms} of the jet 
$\eta$ distribution can be a useful discriminator. The $\left<\eta^2\right>$ 
variable is defined using only the leading six jets. We use $\cal{R}$=0.5 
cone jets for this variable.

We calculate $\left<\eta^2\right>$ by taking the square of the difference
between each jet $\eta$ and the $E_{T}$-weighted mean, $\bar{\eta}$, weighted 
by a factor ${\cal W}(E_T)$. ${\cal W}(E_T)$ depends upon the difference in 
{\sc rms} between $t\bar{t}$ signal ($\sigma_{\eta}^{t\bar{t}}$) and background 
($\sigma_{\eta}^{\rm bkg}$), and is larger at those $E_T$ values where signal 
and background are expected to differ. The $\left<\eta^2\right>$ parameter 
is given by:
\begin{equation}
{\rm \left<\eta^2\right>} = \frac{ \sum_{j=1}^{6} {\cal{W}}(E_{T_j}) 
\left( \eta_{j} - \bar{\eta} \right)^2 } {\sum_{j=1}^{6} {\cal{W}}(E_{T_j}) },
\end{equation}
\noindent
where
\begin{equation}
{\cal{W}}(E_T) = \frac{ \sigma_{\eta}^{t\bar{t}}(E_T) - \sigma_{\eta}^{\rm bkg}
(E_T) }{ \sigma_{\eta}^{t\bar{t}}(E_T)} \quad {\rm and} \nonumber
\end{equation}
\begin{equation}
\bar{\eta} = 
\frac{1}{H_T} \sum_{j=1}^{N_{\rm jets}} E_{T_{j}}~\eta_{j}. \nonumber
\end{equation}
\noindent
Note that both $\sigma_{\eta}^{t\bar{t}}(E_T)$ and $\sigma_{\eta}^{\rm bkg}
(E_T)$ depend on the $E_T$ of the jets in the $\eta$ distribution.  Jets with 
lower $E_T$ tend to be at larger values of $|\eta|$, and consequently 
$\sigma_{\eta}$ decreases with increasing $E_T$. The QCD multijet background 
has a broader distribution in the $\left<\eta^2\right>$ variable than the 
$t\bar{t}$ signal.
\end{enumerate}

\begin{figure}
\centerline{\psfig{figure=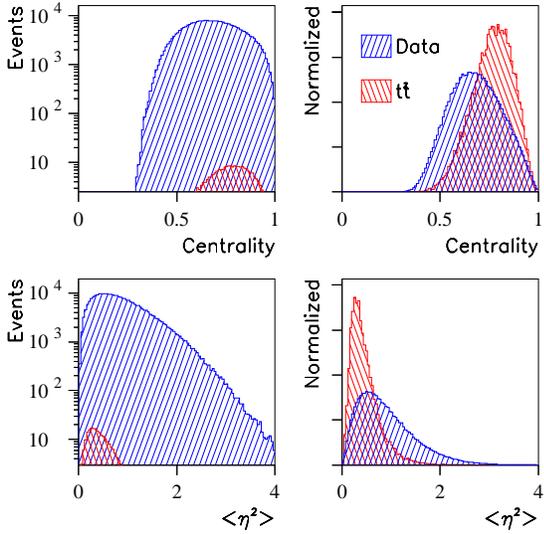,width=3in}}
\caption[The centrality and etc. distributions.]
{\small The centrality and $\left<\eta^2\right>$ distributions for data 
(predominantly background) and for {\sc herwig} $t\bar{t}$ events. Each 
distribution is normalized to the expected number of events (left) and 
to unity (right).}
\label{ccccen}
\end{figure}

The ${\cal C}$ and $\left<\eta^2\right>$ distributions are shown 
in Fig.~\ref{ccccen}, for $m_t$ = 175 GeV/$c^2$.

The above ten kinematic variables are employed as inputs to the first neural
network.  The output of this network is an input to the second (and final) 
neural network, whose three other inputs are described in the following section.
%
\section{Event Structure Variables}
\label{otherparams}

In addition to the kinematic and topological characteristics examined in
Sec.~V, there are other differences between the $t\bar{t}$
signal and the QCD multijet background that we will exploit
in extracting the $t\bar{t}$ signal.
\subsection{$p_T$ of tagging muon}

\begin{figure}
\centerline{
\psfig{figure=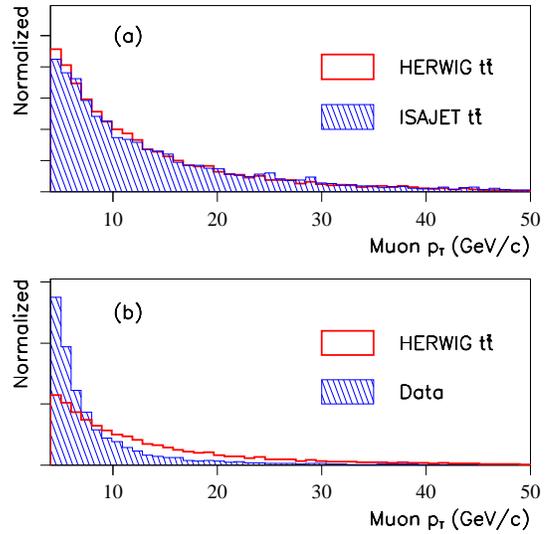,width=3in} }
\caption[PT Dependence]{\small Comparison of muon $p_T$ spectra for
(a) {\sc herwig} and {\sc isajet} $t\bar{t}$ events,
and (b) {\sc herwig} $t\bar{t}$ events and data. These distributions have
been normalized to unity.}
\label{muonpt}
\end{figure}

The $p_T$ of the tagging muon gives further discrimination between $t\bar{t}$ 
signal and QCD background. Not only does the fragmentation of $b$ quarks 
produce higher $p_T$ objects, but the $b$ quark is also more energetic in 
$t\bar{t}$ events than in background.  Thus, the mean muon $p_T$, $p_T^{\mu}$, 
is significantly larger in $t\bar{t}$ events.  Figure~\ref{muonpt} shows the 
muon $p_T$ spectra. Figure~\ref{muonpt}(a) compares the muon $p_T$ in 
{\sc herwig} and {\sc isajet} $t\bar{t}$ events, which shows that the muon 
$p_T$ spectrum is modeled consistently by Monte Carlo. Figure~\ref{muonpt}(b) 
compares {\sc herwig} $t\bar{t}$ events and data (predominantly background).
These results show that the $p_T$ of the muon can serve as a useful tool in
differentiating between signal and background.

%
\subsection{Widths of jets}

At the simplest level, each $t(\bar{t})$ quark decays into a $b(\bar{b})$ 
quark and a $W^{+}(W^{-})$ boson, with each  $W$ boson decaying into light 
quarks. Barring extra gluon bremsstrahlung, this represents six quark-jets 
in the final state.  The average jet multiplicity for {\sc herwig} $t\bar{t}$ 
events ($m_t$=175 GeV/$c^2$) using our selection criteria is 6.9, implying
that the contribution from gluons is relatively small. Conversely, jets in 
the QCD multijet background originate predominantly from gluon radiation.  
Although gluon splitting can take place, producing both quark and gluon jets, 
it is expected that gluons dominate QCD multijet production.

QCD predicts substantial differences between quark jets and gluon jets and,
in fact, observed differences in quark and gluon jet widths  have
been reported by experiments at the KEK $e^+e^-$ collider (TRISTAN)\cite{KEK}
and the CERN $e^+e^-$ collider (LEP)\cite{LEP1}. Parton shower Monte Carlos 
such as {\sc herwig} have been shown to reproduce the widths observed
in data~\cite{LEP1}, although {\sc herwig} has been found to
slightly underestimate jet widths at the Fermilab Tevatron~\cite{ABBOTT}.
We found that by applying a correction of 3\% to the widths, {\sc herwig} QCD
Monte Carlo reproduces the observed distributions in the width of the jets.  
Further studies have shown that the kinematic distributions of the multijet 
background are also well modeled using {\sc herwig}.  We have therefore chosen 
{\sc herwig} as the generator for studying jet widths, with a 3\% correction 
applied to the widths of each jet.

\begin{figure}
\centerline{
\psfig{figure=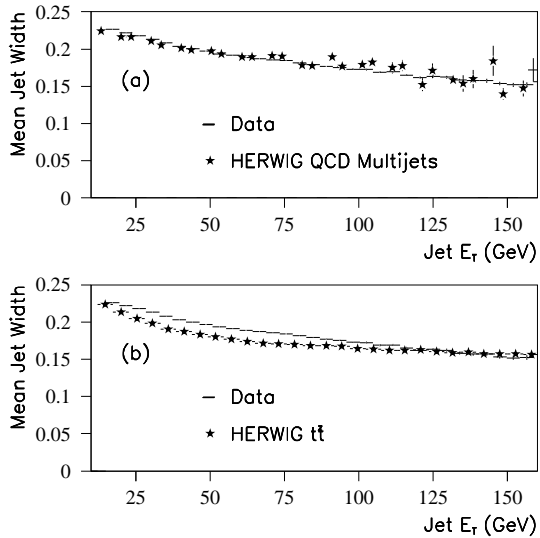,width=3in} }
\caption[ET Dependence] {The mean width of 0.5 cone jets versus their $E_T$ for
(a) data (bars) and {\sc herwig} QCD (stars), and (b) data (bars) and 
{\sc herwig} $t\bar{t}$ (stars). }
\label{TOPBAC}
\end{figure}

Figure~\ref{TOPBAC}(a) shows the mean width of 0.5
cone jets versus jet $E_T$ for multijet data and {\sc herwig} QCD and 
Fig.~\ref{TOPBAC}(b) compares the data to {\sc herwig} $t\bar{t}$.
Here, the jet width is:
\begin{equation}
\sigma_{\rm jet} = \sqrt{\sigma_{\eta}^2+\sigma_{\phi}^2},
\end{equation}
\noindent
where $\sigma_{\eta}$ and $\sigma_{\phi}$ are the transverse energy weighted 
{\sc rms} widths in $\eta$ and $\phi$, respectively, and are calculated using 
the ($\eta$,$\phi$) positions of each calorimeter bin 
(0.1 $\times$ 0.1 in $\Delta\eta$ $\times$ $\Delta\phi$) weighted by the 
transverse energy in that bin. In order to account for the broadening of jets 
from additional minimum bias interactions which could overlap an event,
corrections were applied to the widths of each jet in the event.  These 
corrections were typically a few percent, and depended, among other factors, 
upon the instantaneous luminosity during that event. These corrections were 
determined by assuming that the energy coming from minimum bias interactions
was uniformly distributed in $\Delta\eta$ and $\Delta\phi$, and therefore the 
measured {\sc rms} of a jet was the sum in quadrature of its true {\sc rms} 
and the {\sc rms} of a uniform distribution.

It is clear from Fig.~\ref{TOPBAC}(a) that {\sc herwig} QCD describes the 
widths observed in the data, and the {\sc herwig} $t\bar{t}$ has significantly 
narrower jets. This suggests that the difference may be due to the different 
mix of gluons and quarks in the two processes.

For Monte Carlo it is possible to match initial state quarks to final state 
reconstructed jets because the {\sc herwig} $t\bar{t}$ events are relatively 
simple. The mapping between quarks and jets requires a tight match in $\Delta 
\cal{R}$ between the initial quark and the jet, as well as a reasonable match 
in energy. The following criteria were employed to define Monte Carlo 
``quark-like jets'':

\begin{itemize}
\item Good quality 0.5 cone jet, reconstructed without merging (not formed 
 from two or more adjacent jets) and with $|\eta|~\leq$~2.5, 
\item Distance between initial quark and its reconstructed jet to be $\Delta 
 \cal{R}$~$\leq$~0.05,
\item The difference in energy between the quark and the jet 
 $\Delta E$~$\leq$~$\sqrt{E_{\rm quark}}$ ($E$ in GeV).
\end{itemize}

\begin{figure}
\centerline{
\psfig{figure=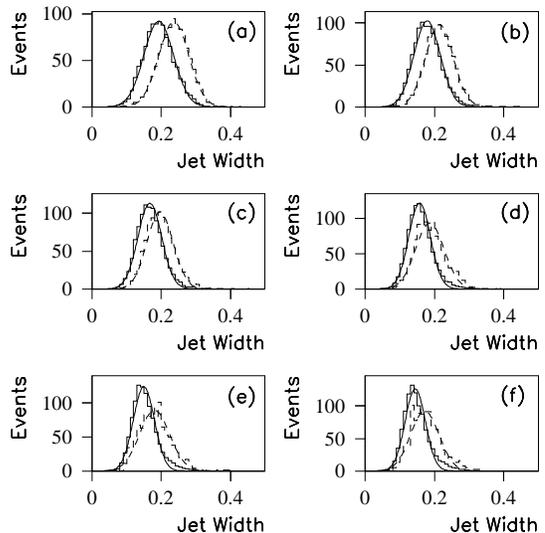,width=3in} }
\caption[ET Dependence] {Distributions in jet {\sc rms} width, 
$\sigma_{\rm jet}$, for {\sc herwig} quark-like jets (solid) and the gluon-like
jets (dashed) for (a) $5 < E_T < 25$ GeV, (b) $20 < E_T < 40$ GeV, 
(c) $35 < E_T < 55$ GeV, (d) $50 < E_T < 70$ GeV, (e) $65 < E_T < 85$ GeV, 
and (f) $80 < E_T < 100$ GeV. These distributions were normalized to have 
equal numbers (1000) of events.}
\label{QMATCH2}
\end{figure}

Monte Carlo ``gluon-like jets'' were defined to be good quality jets, without 
merging, but where the separation distance to the nearest quark was $\Delta 
\cal{R}$~$\geq$~1.  Imposing these criteria, the distributions in the jet 
{\sc rms} widths are shown in Fig.~\ref{QMATCH2}. To guide the eye, Gaussian 
fits have been superimposed on the distributions. With these definitions, it 
appears that gluon-like jets are 20-30\% wider than quark-like jets.

Figure~\ref{QMATCH2} suggests that the jet {\sc rms} distributions for these 
definitions of quark/gluon jets can be approximated by Gaussians.  A Fisher 
discriminant can be used to differentiate statistically between any two such 
distributions.  We defined a Fisher discriminant, ${\cal F}_{\rm jet}$, in 
terms of the individual jet width $\sigma_{\rm jet}$ and the width expected 
for gluon-like ($\sigma_{\rm gluon}$) and quark-like ($\sigma_{\rm quark}$) 
jets, as follows:
\noindent

\begin{equation}
{\cal F}_{\rm jet} = \frac{(\sigma_{\rm jet}-\sigma_{\rm quark}(E_T))^2}
{\sigma^{2}_{\rm quark}(E_T)} - \frac{(\sigma_{\rm jet}-
\sigma_{\rm gluon}(E_T))^2}{\sigma^{2}_{\rm gluon}(E_T)}
\end{equation}

\begin{figure}
\centerline{
\psfig{figure=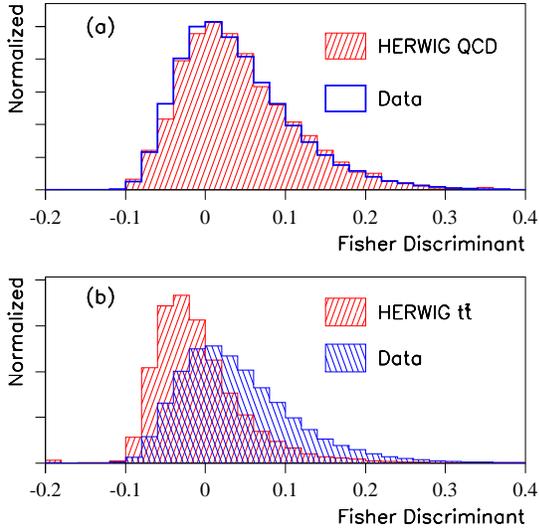,width=3in} }
\caption[ET Dependence] {Distributions of ${\cal F}$ for, (a) data 
(predominantly background) and {\sc herwig} QCD, and (b) data and {\sc herwig} 
$t\bar{t}$ events.}
\label{FISHER}
\end{figure}

We used this single parameter to characterize the quark-like or gluon-like 
essence of a jet.  This discriminant is summed over the four unmerged jets 
with the smallest values of ${\cal F}_{\rm jet}$ in an event to form a 
variable ${\cal F}$ which reflects whether the event is more $t\bar{t}$-like 
(signal) or more QCD-like (background).  Summing only over the four smallest 
values of ${\cal F}_{\rm jet}$ (most quark-like jets), according to Monte 
Carlo, optimizes the discrimination.  This summed discriminant, ${\cal F}$, 
will be used in our search for $t\bar{t}$ signal in the all-jets channel.
The distributions of ${\cal F}$ are shown in Fig.~\ref{FISHER}.  It is known 
that jet widths are not as well modeled in {\sc isajet}~\cite{abbotthesis}, 
and we have, therefore, based this discriminant only on the {\sc herwig} 
generator. Figure~\ref{FISHER}(a) shows ${\cal F}$ for data and {\sc herwig} 
QCD, and Fig.~\ref{FISHER}(b) shows ${\cal F}$ for data and {\sc herwig} 
$t\bar{t} \rightarrow$ all-jets. Comparison shows that the jets in data are 
significantly wider, and are more consistent with {\sc herwig} QCD than with 
{\sc herwig} $t\bar{t}$.

%
\subsection{Mass likelihood parameter}

A mass likelihood variable, ${\cal M}$, defined below,
provides good discrimination between signal and background by requiring 
two jet pairs that are consistent with the $W$ boson mass, and two $W$ + jet
pairs that are consistent with a single top quark mass of {\it any} value.
Since there are no high-$p_T$ leptons in the all-jets channel, and hence no 
high-$p_T$ neutrinos, the event is in principle fully reconstructible.  The 
presence of two $W$ bosons in $t\bar{t}$ events provides significant rejection 
against QCD background.  A further requirement that the two reconstructed top 
quarks have equal masses provides some additional discrimination.
${\cal M}$ is defined as a $\chi^2$-like object:

\begin{equation}
{\cal M} = 
\frac{(M_{W_1}-M_{W})^2}{\sigma_W^2} + \frac{(M_{W_2}-M_{W})^2}{\sigma_W^2}
 + \frac{(m_{t_1}-m_{t_2})^2}{\sigma_t^2},
\end{equation}
\noindent
where $M_{W_1}$ ($M_{W_2}$) is the mass of the two ${\cal{R}}$=0.5 cone jets 
corresponding to the $W$ boson from the first (second) top quark, of mass 
$m_{t_1}$ ($m_{t_2}$). The parameters $M_{W}$, $\sigma_W$ and $\sigma_t$ were 
fixed at 80, 16 and 62 GeV/$c^2$, respectively.  The last two values 
approximate the full widths of the two distributions, and taking them to 
be constant simplifies the calculation.

The ${\cal M}$ variable is calculated by looping over combinations of jets, 
and assigning all jets with $|\eta|~\leq$ 2.5 to one of the $W$ bosons or $b$ 
quarks from the two top quark decays. The smallest value of ${\cal M}$ is 
selected as the discriminator. To reduce the number of combinations, two jets 
are assigned to each $W$ boson, and one to the $b$ quark from one of the two 
top quarks. Jets from the $W$ boson are required to have $E_T > 10$ GeV, while 
those from the $b$ quark must have $E_T > 15$ GeV. All remaining jets are 
assigned to the $b$ quark from the second top quark.  To keep $b$-tagged events
on the same footing as untagged events, no {\it a priori} assignment is made 
between tagged jets and $b$ quarks. Since in the top quark rest frame the $W$ 
boson and the $b$ quark have equal momenta, the $E_T$ of $W$ bosons and 
$b$-jets are more similar than for QCD background.  The following criterion 
helps further reduce combinatorics:

\begin{itemize}
\item $E_{T(W_1)}~+~E_{T(W_2)}~\leq~0.65~H_T$,
\end{itemize}
\noindent
where $E_{T(W_1)}$ ($E_{T(W_2)}$) is the $E_T$ from the vector sum of two jet 
momenta assigned to the $W$ boson from the first (second) top quark.  Although 
there are other possible algorithms for assigning jets to the two top quarks, 
the discrimination in the ${\cal M}$ variable is not very sensitive to the 
choice of reasonable algorithms.

\begin{figure}
\centerline{\psfig{figure=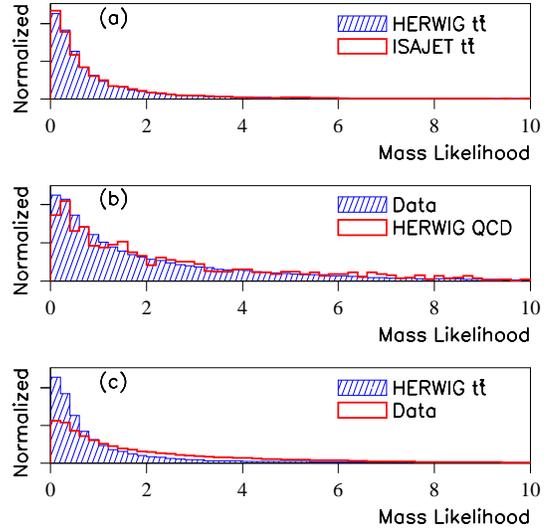,width=3in}}
\caption{\small Distribution in mass likelihood parameter for (a) {\sc herwig} 
and {\sc isajet} $t\bar{t}$ events, (b) {\sc herwig} QCD and data, and (c) 
{\sc herwig} $t\bar{t}$ events and data. These distributions were normalized 
to unity.}
\label{MLL}
\end{figure}

The distributions in the ${\cal M}$ variable are shown in Fig.~\ref{MLL}.  
Figure~\ref{MLL}(a) compares the ${\cal M}$ variable in {\sc herwig} and 
{\sc isajet} $t\bar{t}$ events ($m_t$=175 GeV/$c^2$). Figure~\ref{MLL}(b) 
compares {\sc herwig} QCD and the data (predominantly background). 
Figure~\ref{MLL}(c) compares {\sc herwig} $t\bar{t}$ events and data.
These plots show that this variable is modeled consistently by the two 
$t\bar{t}$ Monte Carlos, that {\sc herwig} QCD models the background well, 
and that ${\cal M}$ is useful in discriminating between signal and background.

%
\subsection{Correlations between parameters}

A summary of the 13 parameters used in this analysis is given in 
Table~\ref{var13list}. The first ten parameters are simple kinematic variables,
and are correlated. To quantify the degree of correlation between any two 
variables $x$ and $y$, we define a linear correlation coefficient, $r$ 
as\cite{bevington}:

\begin{table}
\caption[Top to all-jets analysis variables.]
{\small The 13 variables used in the neural network analysis, the jet cone 
size employed and the $t\bar{t}$ event characteristic upon which it 
discriminates are given.}
\begin{small}
\begin{center}
\begin{tabular}{c|c|c|c}
 Variable &  Description & Cone & Characteristic \\
\hline \hline
$H_T$                 &  Total transverse energy    & 0.3      &    Energy \\
\hline
$\sqrt{\hat{s}}$      &  Total $t\bar{t}$           & 0.3      &    Energy \\
                      &  center-of-mass energy      &          &    Energy \\
\hline
$E_{T_{1}}$/$H_T$     &  Leading jet transverse     & 0.5/0.3  &    Energy \\
                      &  energy fraction            &          &           \\
\hline
$H_T^{3j}$            &  Transverse energy of       & 0.3      &  Radiation \\
                      &  non-leading jets           &          &            \\
\hline
$N_{\rm jets}^A$      &  Weighted number of jets    & 0.3      &  Radiation \\
\hline
$E_{T_{5,6}}$         &  $E_T$ of 5th and           & 0.3      &  Radiation \\
                      &  6th jets                   &          &            \\
\hline
${\cal{A}}$           &  Aplanarity                 & 0.3       &  Topology \\
\hline
${\cal{S}}$           &  Sphericity                 & 0.3       &  Topology \\
\hline
${\cal{C}}$           &  Centrality                 & 0.3       &  Topology \\
\hline
\etarms               &  Rapidity distribution      & 0.5       &  Topology \\
\hline
\hline
$p_T^{\mu}$           &  $p_T$ of tagging muon      &  -        &  Event    \\
                      &                             &           &  Structure \\
\hline
${\cal F}$            &  Fisher discriminant        & 0.5       &  Event    \\ 
                      &  based on jet widths        &           &  Structure \\
\hline
${\cal M}$            &  Mass likelihood            & 0.5       &  Event    \\
                      &                             &           &  Structure \\
\hline
\end{tabular}
\end{center}
\end{small}
\label{var13list}
\end{table}

\begin{equation}
r = { {\displaystyle N\sum }
                    x_iy_i - \sum x_i \sum y_i
           \over
                \left[ {\displaystyle N }
                      \sum x_i^2 - (\sum x_i)^2 \right]^{1/2}
             \left[ N \sum y_i^2 - (\sum y_i)^2 \right]^{1/2}
          }.
\end{equation}

The value of $r$ ranges from 0, when there is no correlation, to $\pm$1, when 
there is complete correlation or anticorrelation. Table~\ref{correlation} 
shows the average correlations among 13 parameters defined in Sec.~V and 
Sec.~VI for data. These are average correlation coefficients; local 
correlations can vary significantly, depending upon the region of multivariate 
space. Note that the parameters $p_T^{\mu}$, ${\cal F}$, and ${\cal M}$
have relatively small correlations with the other kinematic parameters.

\begin{table*}
\caption{\small Average correlations among the 13 parameters for data.}
\begin{small}
\begin{center}
\begin{tabular}{c|rrrrrrrrrrrrr}
& $H_T$ & $\sqrt{\hat{s}}$ & $E_{T_{1}}/H_T$ & $H_T^{3j}$ & $N_{\rm jets}^A$ & 
$E_{T_{5,6}}$ & \Aplan & \Spher & ${\cal{C}}$ & \etarms & $p_T^{\mu}$ & ${\cal 
F}$  & ${\cal M}$ \\
\hline
$H_T$             & 1  & 0.80 & --0.14 &   0.71 &   0.76 &   0.39 &   0.01 
&   0.   &   0.17 & --0.31 &   0.04 & --0.04 &   0.05 \\
$\sqrt{\hat{s}}$  &    & 1    & --0.20 &   0.64 &   0.64 &   0.36 & --0.16 
& --0.25 & --0.32 &   0.14 &   0.01 & --0.08 &   0.05 \\
$E_{T_{1}}/H_T$   &    &      &   1    & --0.54 & --0.36 & --0.37 & --0.34 
& --0.23 &   0.07 &   0.14 & --0.02 &   0.23 &   0.30 \\
$H_T^{3j}$        &    &      &        &   1    &   0.76 &   0.71 &   0.25 
&   0.15 &   0.05 & --0.25 &   0.04 & --0.02 & --0.10 \\
$N_{\rm jets}^A$  &    &      &        &        &   1    &   0.44 &   0.12 
&   0.09 &   0.09 & --0.27 &   0.04 & --0.05 & --0.04 \\
$E_{T_{5,6}}$     &    &      &        &        &        &   1    &   0.21 
&   0.12 &   0.02 &   0.02 &   0.03 & --0.03 & --0.10 \\
\Aplan            &    &      &        &        &        &        &   1    
&   0.58 &   0.26 & --0.30 &   0.04 & --0.07 & --0.16 \\
\Spher            &    &      &        &        &        &        &        
&   1    &   0.37 & --0.40 &   0.03 & --0.04 & --0.14 \\
${\cal{C}}$       &    &      &        &        &        &        &        
&        &   1    & --0.59 &   0.05 &   0.06 &   0.   \\
\etarms           &    &      &        &        &        &        &        
&        &        &   1    & --0.05 & --0.07 &   0.03 \\
$p_T^{\mu}$       &    &      &        &        &        &        &        
&        &        &        &   1    & --0.01 &   0.   \\
${\cal F}$        &    &      &        &        &        &        &        
&        &        &        &        &   1    &   0.10 \\
${\cal M}$        &    &      &        &        &        &        &        
&        &        &        &        &        &   1    \\
\end{tabular}
\end{center}
\end{small}
\label{correlation}
\end{table*}
%
\section{Analysis}
\subsection{Event selection criteria}

Before proceeding further with the analysis, basic quality criteria were 
applied to the data and to Monte Carlo events:

\begin{itemize}
\item{\bf isolated leptons:}  Events containing an isolated electron or 
 muon \cite{ouroldprd,meenajim} were rejected. This ensured that our event 
 sample was orthogonal to those used in the $t\bar{t}$ analyses in other decay 
 channels.
\item{\bf $H_T^{3j}\geq$ 25 GeV:} Removed QCD $2 \rightarrow 2$ events with 
 little additional jet activity.
\item{\bf number of jets:}  Events with fewer than six $\cal{R}$=0.3 cone jets 
 or more than eight $\cal{R}$=0.5 cone jets were rejected.
  \begin{itemize}
    \item  By eliminating events with fewer than six $\cal{R}$=0.3 cone jets, 
    the signal-to-background ratio is improved.  Only 14\% of the signal is 
    lost, while 36\% of the background is rejected.  (The $E_T$ of the sixth 
    jet is required in the calculation of several variables.)
    \item Removal of events with more than eight $\cal{R}$=0.5 cone jets also 
    improves signal-to-background, rejecting 13\% of the background and only 
    5\% of the signal.  The calculation of the ${\cal M}$ variable and Fisher 
    discriminant are thereby improved because of the reduction in the number 
    of jet combinations.
  \end{itemize}
\end{itemize}

Of the roughly 600,000 events passing our initial criteria (see Table II),
approximately 280,000 events survive these selection requirements.

\subsection{Muon tagging}

The direct branching fraction of a $b$ quark into a muon plus anything
is 10.7 $\pm$ 0.5\%\cite{PDG}. However, when all contributions from decays of 
$b$ and $c$ quarks from the two top quarks are considered, and with a muon 
acceptance of about 50\%, approximately 20\% of the events in the $t\bar{t}$ 
$\rightarrow$ all-jets mode are expected to yield at least one muon. Muons 
in QCD background processes arise mainly from gluon splitting into $c \bar{c}$ 
or $b \bar{b}$ pairs, but intrinsic $c \bar{c}$ and $b \bar{b}$ production 
as well as in-flight pion and kaon decays within jets also contribute. These 
sources occur in only a small fraction of the events, and therefore only a few 
percent of the QCD multijet background events will have a muon 
tag\cite{ouroldprd}.

To take advantage of the difference in the muon tag rate and enhance the 
$t\bar{t}$ signal, our analysis requires the presence of at least one muon 
near a jet in every event (``$b$-tagging''). This also provides a means of 
estimating the background in a given data sample, which can be determined 
purely from data. The $b$-tagging requirement should give nearly a factor 
of ten improvement in signal/background\cite{ouroldprd}.

Procedures for tagging jets with muons were defined after extensive Monte Carlo 
studies of $t\bar{t}$ production in lepton+jets final states \cite{ouroldprd}.
The requirements used to select such muon tags are:

\begin{itemize}
\item The presence of a
fully reconstructed muon track in the central region ($|\eta|<$1.0). 
This restriction does not have much impact on the acceptance of muons 
from $b$ quark jets from $t\bar{t}$ decay because these $b$ quarks tend 
to be produced mainly at central rapidities.

\item The track must be flagged as a high-quality muon. This quality is
based on a $\chi^2$ fit to the track in both the bend and non-bend views of 
the muon system \cite{muonstudy}.

\item The signal from the calorimeter in the road defined by the track must 
be consistent with the  passage of a minimum ionizing particle. The signal 
is measured by energy deposited in the calorimeter cells along the track.

\item Because the $p_T$ spectrum of muons from pion and kaon decays is
softer than from heavy quarks, an overall $p_T >$ 4.0 GeV/$c$
cutoff is imposed to enhance the signal from heavy quarks.  Imposing
this cutoff has limited impact on the $t\bar{t}$ acceptance, since
the muon energy must be greater than about 3.5 GeV in order to penetrate
the material of the calorimeter and the iron toroid at $\eta$=0.

\item The muon must be reconstructed near a jet that has $|\eta|<$1.0 and
$E_T >$ 10 GeV.  The distance $\Delta {\cal R}_{\mu}$ in $\eta$-$\phi$ space
between the muon and the jet axis must be less than 0.5.
\end{itemize}

If a muon satisfies the above conditions, the jet associated with
the muon is defined as a $b$-tagged jet, and the muon is called a tag.
Of the roughly 280,000 events which survived the initial selection criteria,
3853 have at least one $b$-tagged jet.

%
\subsection{Muon tagging rates}
\label{tagratefun}

The probability of tagging QCD background events containing several jets is 
observed to be just the sum of the probabilities of tagging individual 
jets\cite{ouroldprd}, and is approximately independent of the nature of the 
rest of the event. The muon tagging rate is therefore defined in terms of 
probability per jet rather than per event. We define the muon tagging rate as 
the ratio of tagged to untagged jets, allowing us to multiply this function by 
the number of untagged events to obtain an estimate of the tagged background.

\begin{figure}
\centerline{\psfig{figure=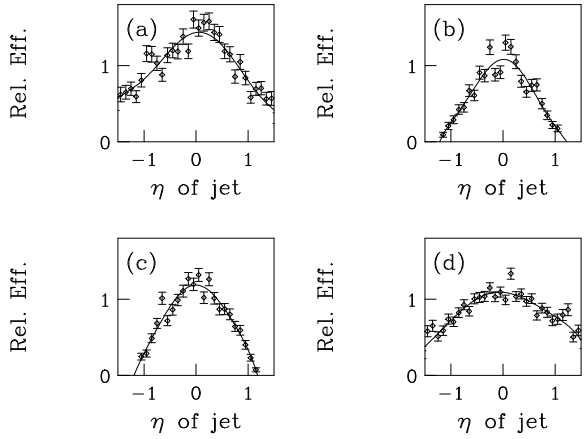,width=3in}}
\caption[The relative muon detection efficiency for different run ranges.]
{\small The relative muon detection efficiency as a function of the $\eta$ of 
the jet, for different ranges of runs: (a) $N_{\rm run}<$70000, (b) 
70000$\leq N_{\rm run}<$84000, (c) 84000$\leq N_{\rm run}<$89000, and (d) 
$N_{\rm run}\geq$89000. The curves represent the results of polynomial fits.}
\label{centag}
\end{figure}

Initially, the tagging rate was modeled only as a function of jet 
$E_T$\cite{d0obs,ouroldprd}. However, it was found subsequently that there 
was an $\eta$-dependence to the muon tag rate which depended on the date of 
the run. This was traced to the fact that the muon chambers experienced 
radiation damage, and required that some of the wires be cleaned during the 
run. Figure~\ref{centag} shows the relative muon detection efficiency as a 
function of the $\eta$ of the jet for different ranges of runs. 
Figures~\ref{centag}(a)--\ref{centag}(c) correspond to the time before 
the cleaning and Fig.~\ref{centag}(d) to that after the cleaning 
($N_{\rm run}~\geq$ 89000). These plots illustrate the need to account 
for the dependence on $\eta$ and run number when performing estimates of 
tagging rates.

To address this problem, the tag rate for background, 
$P_{\rm tag}$($E_T$,$\eta$,$N_{\rm run}$), was parameterized as a function 
of jet $E_T$, jet $\eta$, and the run number, $N_{\rm run}$, and was assumed to
factorize:

\begin{equation}
P_{\rm tag}(E_T,\eta,N_{\rm run}) = f(E_T) \cdot \epsilon(\eta,N_{\rm run}),
\label{tagrate}
\end{equation}
where $f(E_T)$ is the relative probability that a jet of given $E_T$ has a 
muon tag, and $\epsilon(\eta,N_{\rm run})$ is the relative muon detection 
efficiency. The functions $f(E_T)$ and $\epsilon(\eta,N_{\rm run})$ are not 
normalized individually, but it is the product of the two which is normalized.

Besides the differences in chamber efficiency caused by the deterioration and 
cleaning of wires, there were also changes in the gas mixtures used in the
muon chambers between the Ia period and Ib (see Table I), and changes in the
high voltage settings, which were implemented at Run 84000. These required two 
additional separations of runs, as shown in Fig.~\ref{centag}. We also found 
a small dependence of the tag rate function on $\sqrt{\hat{s}}$ of the entire 
event, which is described below. 

\begin{figure}
\centerline{\psfig{figure=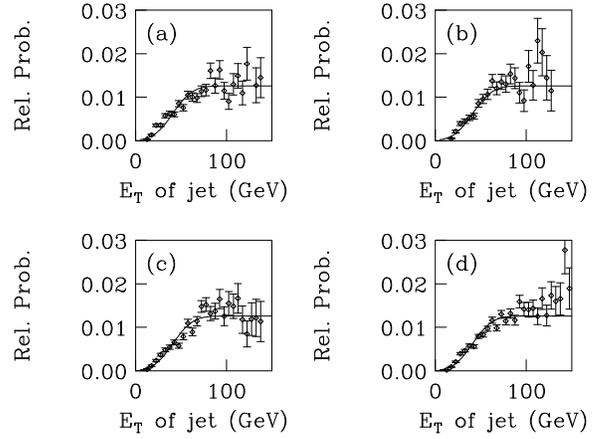,width=3in}}
\caption[The muon tag rate in $E_T$ for different run ranges.]
{\small The relative probability, $f(E_T)$, for central jets as a function of 
the jet $E_T$, for runs in the range: (a) $N_{\rm run}<$70000, (b) 70000$\leq 
N_{\rm run}<$84000, (c) 84000$\leq N_{\rm run}<$89000, and (d) 
$N_{\rm run}\geq$89000.  The curves represent the results of a
common fit using Eq.~\ref{cenmuonfit1}, and saturate at high jet $E_T$.}
\label{tags1}
\end{figure}

The jet $E_T$ factor in the muon tag rate function ($f(E_T)$) is shown in 
Fig.~\ref{tags1}. $f(E_T)$ was parameterized in two ways, which allowed us to 
estimate a systematic error due to the model dependence of this function.  
The first parameterization assumed that $f(E_T)$ saturates at high values of 
jet $E_T$, and was given by the form:
\begin{equation}
f(E_T) = A_0~\aleph(\frac{E_T - E_{T_0}}{\lambda}),
\label{cenmuonfit1}
\end{equation}
\noindent
where $\aleph(x)$ is the normal frequency function ({\it i.e.}, 
$\aleph(x) = \frac{1}{\sqrt{2\pi}}\int_{-\infty}^{x} e^{-z^2/2} dz$), which 
approaches one at high jet $E_T$.  The parameters $A_0, E_{T_0}$, and 
$\lambda$ are obtained from the fits to the observed tag rates,
shown in Fig.~\ref{tags1}.

\begin{figure}
\centerline{\psfig{figure=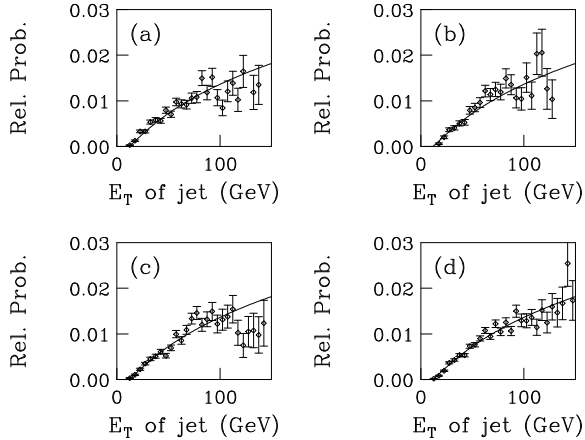,width=3in}}
\caption[The muon tag rate in $E_T$ for different run ranges.]
{\small The relative probability, $f(E_T)$, for central jets as a function of 
the jet $E_T$, for runs in the range: (a) $N_{\rm run}<$70000, (b) 70000$\leq 
N_{\rm run}<$84000, (c) 84000$\leq N_{\rm run}<$89000, and (d) 
$N_{\rm run}\geq$89000.  The curves represent the results of a
common fit using Eq.~\ref{cenmuonfit2}, and do not saturate at high jet $E_T$.}
\label{tags2}
\end{figure}

An alternative parameterization of $f(E_T)$ assumed a polynomial in ln($E_T$), 
and was given by:
\begin{equation}
f(E_T) = a_0 + a_1~{\rm ln}(E_T) + a_2~{\rm ln}^{2}(E_T) 
       + a_3~{\rm ln}^{3}(E_T).
\label{cenmuonfit2}
\end{equation}
\noindent
Here, $f(E_T)$ continues to increase with jet $E_T$, and the constants 
$a_0,~a_1,~a_2$, and $a_3$ are again obtained from fits to the observed 
tagged distributions, shown in Fig.~\ref{tags2}.  The difference in the 
background estimate between Eq.~\ref{cenmuonfit1} and Eq.~\ref{cenmuonfit2} 
is discussed in Sec.~VII.I.  Because the tagging rate in Eq.~\ref{cenmuonfit2} 
continues to grow with increasing jet $E_T$, it gives a slightly larger 
estimate of the background than Eq.~\ref{cenmuonfit1}. Increasing the tag 
rate increases the estimated background, thereby decreasing the signal. 
Both versions of $f(E_T)$ give similar $\chi^2$ fits, but as our Monte Carlo
studies showed that the tag rate continues to slowly increase with jet
$E_T$, even for high $E_T$, we chose equation~\ref{cenmuonfit2} for estimating
the background in this analysis.

\begin{figure}
\centerline{\psfig{figure=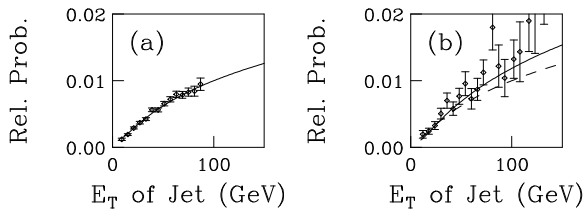,width=3in}}
\caption[The muon tag rate in $E_T$ for different shat ranges.]
{\small The relative probability, $f(E_T)$, for central jets as a function of 
the jet $E_T$, for (a) 200$<\sqrt{\hat{s}}<$300 GeV/$c^2$ and (b) 
400$<\sqrt{\hat{s}}<$500 GeV/$c^2$. Solid curves represent fits using 
Eq.~\ref{cenmuonfit2}, including a dependence on $\sqrt{\hat{s}}$.
The dashed curve represents the fit at 200$<\sqrt{\hat{s}}<$300 GeV/$c^2$.}
\label{tags3}
\end{figure}

Having considered all factors that go into the tag rate function on a 
jet-by-jet basis, we looked for dependence on characteristics of the event 
as a whole. We observed a small additional dependence, most notable in 
variables that are sensitive to the total energy scale of the event.  
Figure~\ref{tags3} shows the muon tag rate in two bins of $\sqrt{\hat{s}}$, 
which reflects the total energy of the partonic collision. The superimposed 
solid curves represent fits to Eq.~\ref{cenmuonfit2}, but where the 
coefficients $a_0,~a_1,~a_2$, and $a_3$ are now second-order polynomials 
in $\sqrt{\hat{s}}$. In Fig.~\ref{tags3}(b), the dashed curve represents 
the fit at 200$<\sqrt{\hat{s}}<$300 GeV/$c^2$, and a small shift in the 
relative tag rate is apparent. This $\sqrt{\hat{s}}$ dependence was included 
through a modification of the principal $E_T$-dependent part of the function, 
$f(E_T)$.

As indicated by Eq.~\ref{tagrate}, the observed tag rate is the product of two 
parts.  Because of this, the fits of Eqs.~\ref{cenmuonfit1} or 
\ref{cenmuonfit2} to the observed tag rate are correlated with the muon 
detection efficiency. To disentangle the two components, the fit used data only
from central rapidities, where the detection efficiency was a weak function of 
$\eta$. The criterion  $\epsilon(\eta,N_{\rm run})/\epsilon(0,N_{\rm run}) 
\geq$0.6 defined the region used in the fit, corresponding to the region where 
the $\eta$-dependence varied least rapidly. Once this initial $f(E_T)$ was 
determined, it was necessary to use it to re-estimate 
$\epsilon(\eta,N_{\rm run})$.  This involved taking the ratio of the number of 
observed tagged jets to the number predicted using the initial $f(E_T)$.  
This ratio, as a function of $\eta$, is plotted in Fig.~\ref{centag} for
different run ranges.  The process of fitting $f(E_T)$ and then re-calculating 
$\epsilon(\eta,N_{\rm run})$ was iterated several times until stable results 
were obtained. The final relative probabilities ($f(E_T)$) are shown in 
Fig.~\ref{tags1} and Fig.~\ref{tags2}, and the final relative efficiency  
is shown in Fig.~\ref{centag}. These are labeled relative 
probabilities/efficiencies because it is not possible to determine the overall 
normalizations of $f(E_T)$ and $\epsilon(\eta,N_{\rm run})$ independently; it 
is their product which is well determined.

Using Eq.~\ref{tagrate}, the number of expected tagged events (from background)
in a given event sample is

\begin{equation}
N_{\rm tag}^{\rm expt} 
 = \sum_{\rm events} ~~\sum_{\rm jets} P_{\rm tag}(E_T,\eta,N_{\rm run}).
\label{tagsum}
\end{equation}
\noindent
In using Eq.~\ref{tagsum} to estimate the tagged background, we assumed that 
this relation remains valid for extrapolation from the background region 
through to the signal region.  These regions will be defined in terms of the 
neural network output, in Sec.~VII.E. This supposes that there is no 
significant correlation between the {\it intrinsic} heavy quark ($c\bar{c}$ 
or $b\bar{b}$) content and the neural network output, apart from any kinematic 
correlation through variation in $E_T$ and $\eta$, as parametrized by 
Eq.~\ref{tagsum}.  Therefore, we attribute any excess of tagged events over 
the background predicted by Eq.~\ref{tagsum} to $t\bar{t}$
production.

%
\subsection{Background modeling}
\label{faketags}

Since the kinematic variables are calculated using the jet energies, they are 
to some extent sensitive to the small shift in energy due to the presence of 
the tagged muon and its associated neutrino. As was described earlier, jets 
are measured through the deposition of energy in the calorimeter, and are not
corrected for the muon's momentum.  The neutrino's energy is, of course, missed
completely, and there is no unique prescription for correcting the jet's energy
for the neutrino.  However, these corrections are typically small 
(of the order of the muon momentum).

Previous analyses~\cite{snydermass} aimed at determining the top quark mass 
have incorporated approximate correction factors for the energies of tagged 
jets.  For our analysis, such corrections are not strictly needed, and as
we argue below, are disfavored due to the correlations they introduce between 
the $E_T$ of the tagged jet and the $p_T$ of the tagging muon. Our procedure 
consists of calculating the muon tag rate function (Eq. 7.1) from jets 
containing muon tags and untagged jets as follows: we denote the distribution 
of untagged jets as a function of $E_T$ by $U(E_T)$, and the distribution of 
the tagged jets by $T(E_T^{\prime}$.   The distribution $U(E_T)$ reflects 
dominantly QCD background. Here, $E_T$ is the transverse energy observed for 
jets with no observable muon, and thus is on average the true jet energy;
$E_T^\prime$ is the observed energy for tagged jets, without corrections, 
and thus is missing the contributions to the progenitor jets due to the 
transverse energy of the muon and neutrino.  We formed the ratio
$T(E_T^{\prime})/U(E_T)$, taking the same numerical values of $E_T^{\prime}$ 
and $E_T$.  This ratiowas then parameterized, as discussed in Sec.~VII.C, to 
give the tag rate function, $P_{\rm tag}(E_T)$. The $E_T$ distribution of QCD 
background events with a tagged jet, $B(E_T)$, for our analysis was then 
obtained using the untagged jet sample $U(E_T)$ from the expression $B(E_T) 
= P_{\rm tag}(E_T) \times U(E_T)$, which, apart from the smoothing applied to 
the tag rate function, is equivalent to $B(E_T)$ = $T(E_T^{\prime})$.

\begin{figure}
\centerline{\psfig{figure=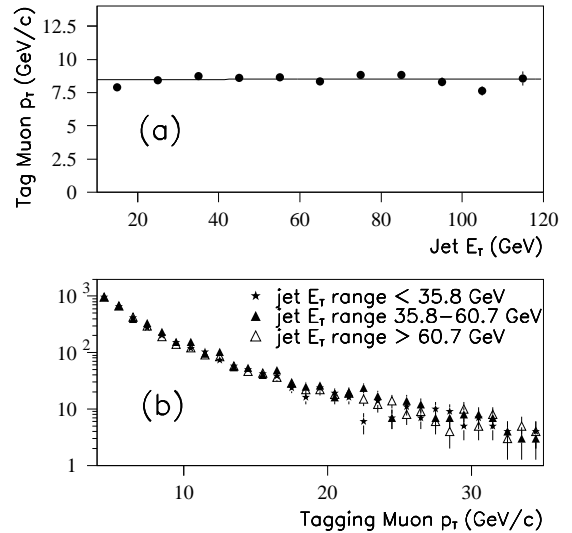,width=3in}}
\caption[PT fits]
{\small (a) Mean muon $p_T$ (dots) versus tagged jet $E_T^{\prime}$ 
and (b) muon $p_T$ distributions for three jet $E_T^{\prime}$ ranges
(chosen to be equally populated) for data events. The line in (a) is the 
average of the points. No correlation is observed between the muon $p_T$ and 
the jet $E_T^{\prime}$, where $E_T^{\prime}$ is the observed energy for tagged 
jets, without corrections (see text).}
\label{etpt}
\end{figure}

\begin{figure}
\centerline{\psfig{figure=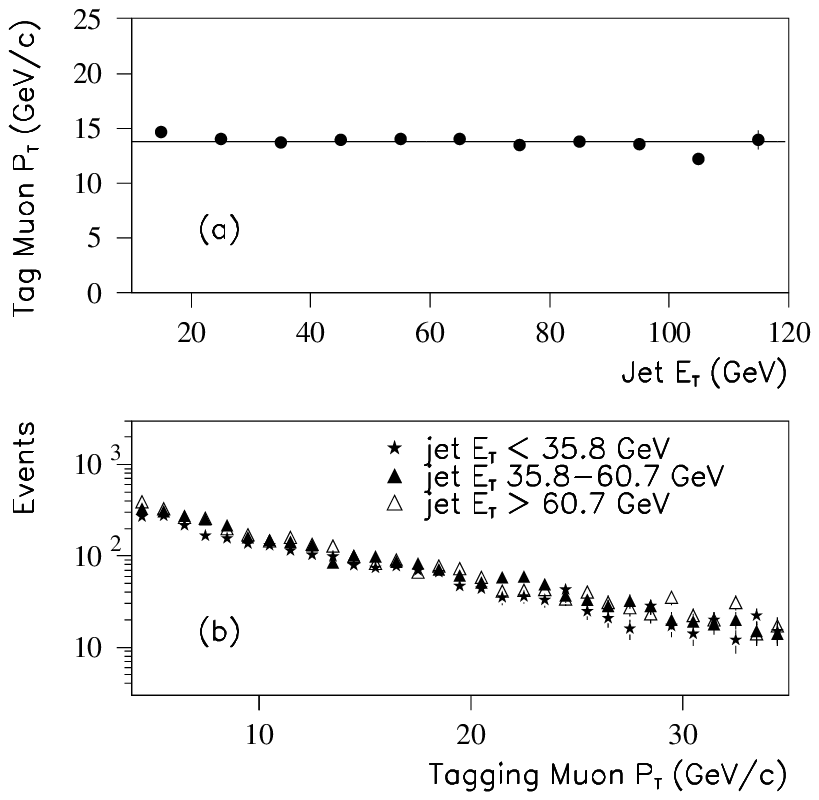,width=3in}}
\caption[PT fits]
{\small (a) Mean muon $p_T$ (dots) versus tagged jet $E_T^{\prime}$ 
and (b) muon $p_T$ distributions for three jet $E_T^{\prime}$ ranges
(chosen to be equally populated) for {\sc herwig} $t\bar{t}$ events.
The line in (a) is the average of the points. No correlation is observed 
between the muon $p_T$ and the jet $E_T^{\prime}$, where $E_T^{\prime}$ is 
the observed energy for tagged jets, without corrections (see text).}
\label{etpt_top}
\end{figure}

Although there is no {\rm a priori} advantage to using uncorrected $E_T^\prime$
instead of corrected $E_T$ for the tagged jets, it does simplify the background
calculation for the neural network analyses.  Our studies show that the $p_T$ 
of the muon is uncorrelated with $E_T^\prime$, but not with $E_T$. This is 
illustrated in Fig.~\ref{etpt}(a), which shows the mean muon $p_T$ as a 
function of the tagged jet $E_T^\prime$ for data.  A fit to a straight line 
gives a slope consistent with zero.  Figure~\ref{etpt}(b) shows muon $p_T$ 
distributions for three distinct ranges of tagged jet $E_T^\prime$ (chosen to 
be equally populated); they are indistinguishable. Similar plots are shown in 
Fig.~\ref{etpt_top} for {\sc herwig} $t\bar{t}$ events. Again, no significant 
correlation between muon $p_T$ and tagged jet $E_T^\prime$ is observed.

\begin{figure}
\centerline{\psfig{figure=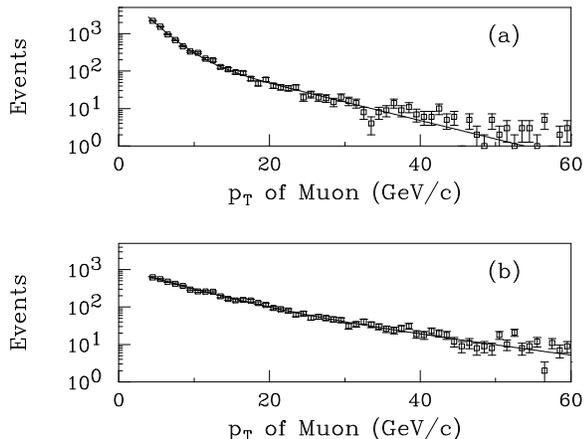,width=3in}}
\caption[PT fits]
{\small Muon $p_T$ distributions for (a) data (predominantly background) and 
(b) {\sc herwig} $t\bar{t}$ events. The smooth curves are from fits to the sum 
of two exponentials. The fact that the curve in (a) is below the points for 
$p_T > 35$ GeV/c does not measurably bias this analysis, because the fraction 
of events in that region is $< 0.6$\%.}
\label{both_pt}
\end{figure}

Since the $p_T$ of the muon is not correlated with the uncorrected jet $E_T$, 
it is largely independent of event kinematics and the probability of finding 
a muon of a given $p_T$ factorizes from the tag rate function.  Tagged 
background events can therefore be generated by adding (``fake'') muons to 
untagged events by assigning a random $p_T$ value from the observed $p_T$ 
spectra. The value of $p_T$ enters into the second neural network and must be 
generated for the modeled background. The $p_T$ distributions for both data 
(predominantly background) and {\sc herwig} $t\bar{t}$ events were fitted 
separately to the sum of two exponentials, and the parameterizations from the 
fits were used in the random generation of muon $p_T$ values for both 
background and signal.  These spectra and the associated fits are shown in 
Fig.~\ref{both_pt}. As discussed above, correcting the jets for muon and 
neutrino $p_T$ would introduce correlations that would complicate the 
application of the tag rate function; we have consequently not applied such 
corrections to the jet energies.


\begin{figure*}
\centerline{\psfig{figure=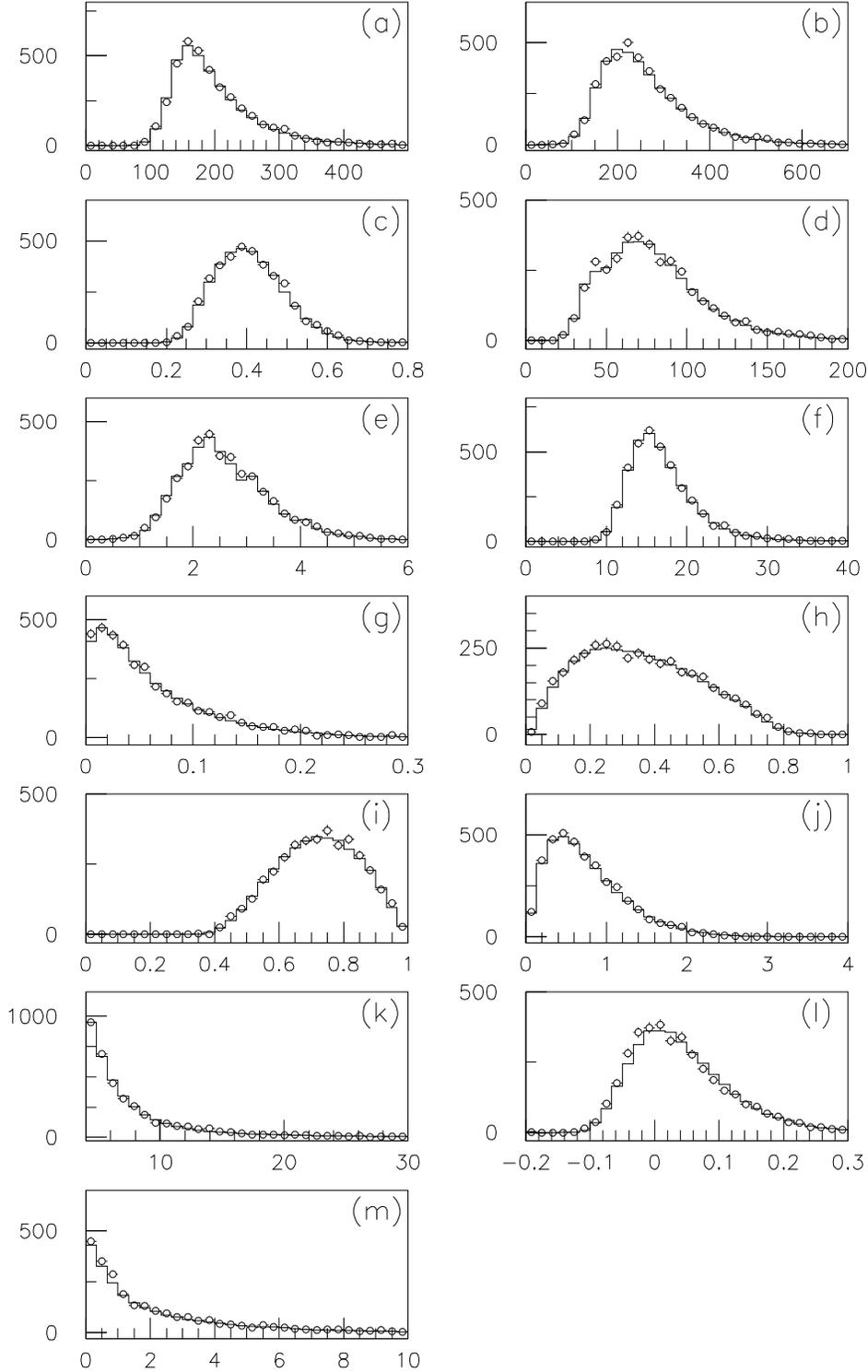,height=8in}}
\caption[Comparisons.]
{\small Comparison of the absolute number of $b$-tagged events expected from 
multijet background (histogram) with the observed (3853) $b$-tagged events in 
data (circles), as a function of each of the 13 variables:  (a) $H_T$ (GeV), 
(b) $\sqrt{\hat{s}}$ (GeV/$c^2$),(c) $E_{T_{1}}$/$H_T$, (d) $H_T^{3j}$ (GeV),
(e) $N_{\rm jets}^A$, (f) $E_{T_{5,6}}$ (GeV), (g) ${\cal{A}}$, (h) ${\cal{S}}$,
(i) ${\cal{C}}$, (j) $\left<\eta^2\right>$, (k) $p_T^{\mu}$ (GeV/$c$), 
(l) ${\cal F}$, and (m) ${\cal M}$.}
\label{muonpred}
\end{figure*}

The procedure used for estimating the number of tagged events expected from 
background can be checked by comparing the distributions of estimated tags to 
those for the observed tags. Figure~\ref{muonpred} shows this comparison for 
the distributions in each of the 13 parameters used in this analysis, for the 
entire multijet tagged data sample.  In these distributions the $t\bar{t}$ 
fraction is negligible, as less than 40 $t\bar{t}$ events are expected. The 
predicted rate, absolutely normalized using Eq.~\ref{tagsum}, is shown for all 
distributions, and consistently reproduces the observed number of tagged events.
The values of $\chi^2$ per degree of freedom for the plots in 
Fig.~\ref{muonpred} are given in Table V.

\begin{table}
 \begin{small}
 \caption{\small $\chi^2$ per degrees of freedom for the plots in 
 Fig.~\ref{muonpred}. For simplicity, only bins with more than ten events were 
 used and only statistical errors were included in the calculations. }
 \begin{center}
 \begin{tabular}{c|c|c}
 Variable &  $\chi^2$ / d.o.f. & Probability of $\chi^2$ \\
\hline
$H_T$                 &  20.1 / 20 & 0.45 \\
$\sqrt{\hat{s}}$      &  25.4 / 25 & 0.44 \\
$E_{T_{1}}$/$H_T$     &  24.1 / 20 & 0.24 \\
$H_T^{3j}$            &  17.5 / 22 & 0.74 \\
$N_{\rm jets}^A$      &  16.9 / 18 & 0.53 \\
$E_{T_{5,6}}$         &  26.7 / 25 & 0.37 \\
${\cal{A}}$           &  15.0 / 23 & 0.89 \\
${\cal{S}}$           &  13.7 / 18 & 0.75 \\
${\cal{C}}$           &  10.0 / 18 & 0.93 \\
$\left<\eta^2\right>$ &  22.0 / 17 & 0.18 \\
$p_T^{\mu}$           &  18.2 / 26 & 0.87 \\
${\cal F}$            &  33.7 / 25 & 0.11 \\
${\cal M}$            &  23.6 / 24 & 0.48 \\
\end{tabular}
 \end{center}
 \end{small}
\end{table}

Once the background sample is generated, these events are treated exactly as 
the tagged sample (the sample used to extract signal). The neural network is 
applied to both sets of events, tagged and modeled background (untagged 
events+``fake-tags''), and the difference between the two represents an excess 
that is attributed to the $t\bar{t}$ signal.  Similarly, ``fake-tags'' are 
applied to the untagged {\sc herwig} $t\bar{t}$ events, and these events are 
used to model the signal.  This effectively increases the statistics of the 
tagged events in the Monte Carlo $t\bar{t}$ sample.

A correction for the small contamination of the background sample due
to $t\bar{t}$ events is made (see Sec.~VII.I).

%
%
\subsection{Neural network analysis}
\label{neuralnetsection}

Artificial neural networks constitute a powerful extension of conventional 
methods of multidimensional data analysis \cite{neural}, and are well suited 
to our search because they handle information from a large number of inputs and
can account for nonlinear correlations between inputs. A neural net is a 
multivariate discriminant. Its construction typically consists of input nodes, 
output(s), and intermediary ``hidden nodes''. The connection between any two 
nodes is governed by a sigmoidal function which is characterized by a 
``weight'' and ``threshold''.  The neural network is ``trained'' by
setting weights and thresholds of the nodes through an optimization algorithm.

The output of the neural network is simply a mapping between the 
multidimensional space described by our kinematic input variables and a 
one-dimensional output space. Setting a threshold on the output of the neural 
network corresponds to a set of hypersurface cuts in multidimensional input 
space. Consequently, the neural network output may be employed to discriminate 
between signal and background as long as the following conditions are observed:

\begin{itemize}
\item The neural network is trained on event samples that are independent of 
 the sample used for the measurement.
\item There is a reliable method for determining the background level for a 
 given value of neural network output.
\end{itemize}
\noindent

Independence of the training sample and the sample used to extract the 
$t\bar{t}$ signal is maintained by considering only $b$-tagged events in the 
final extraction of a signal for $t\bar{t}$ production. Events that did not 
have a $b$-tagged jet are used for training and for defining the background 
sample.

In order to simulate the background, untagged events were made to resemble
tagged events by adding muon tags to one of the jets in the event. With such 
``fake'' muons, these events were taken to represent the background.  The 
prescription for adding these muons to the untagged jets was described in 
Sec.~VII.D.  A subset of these events was used to train the neural network 
response to background.

The set of 13 parameters (see Table~\ref{var13list}) was used as the set of 
input nodes in training the neural network. Because training time increases 
markedly and quality of convergence decreases with the number of input nodes 
and hidden layers, the problem was simplified by first training a neural 
network using the first ten kinematic variables.  These variables tended to 
be more highly correlated than the remaining three (see Sec.~VI). Based on 
studies using our training samples, we chose to have 20 hidden nodes and one 
network output, and used the back-propagation learning algorithm in 
{\sc jetnet} \cite{jetnet}. The output of this neural network and the remaining
three parameters were used as inputs to a second neural network.  Here, we 
chose eight hidden nodes and one network output.

\begin{figure}
\centerline{\psfig{figure=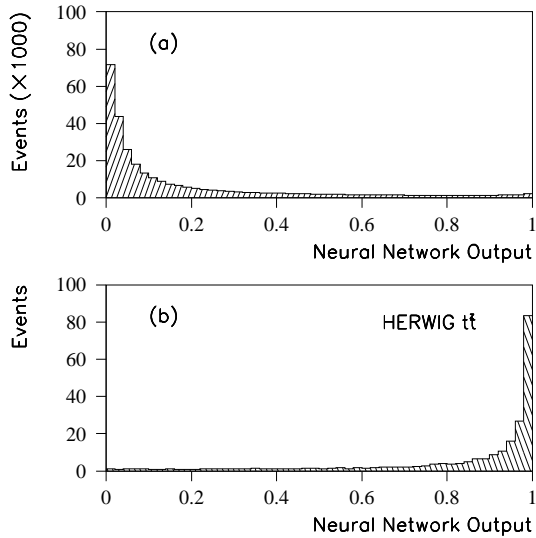,width=3in}}
\caption[]{\small Initial training of the neural network (NN$_0$). The network 
 output is shown for (a) data, and (b) {\sc herwig} $t\bar{t}$ Monte Carlo 
 for $m_t$=180 GeV/$c^2$.}
\label{sig0820}
\end{figure}

Events used to train the two neural networks were selected as follows. A 
simpler initial network (NN$_0$), using a subset of seven kinematic parameters 
(excluding $E_{T_{1}}$/$H_T$, $E_{T_{5,6}}$, and $\left<\eta^2\right>$), was 
trained using all events.  The output of this network, for both data and 
{\sc herwig} $t\bar{t}$ Monte Carlo, is shown in Fig.~\ref{sig0820}.  
Figure~\ref{sig0820} shows that the $t\bar{t}$ signal tends to peak at values 
of neural network output near 1 (the ``signal region''), whereas the background
events peak near 0 (the ``background region'').  For the final training samples,
we selected data and $t\bar{t}$ Monte Carlo events having NN$_0 > $ 0.3. This 
neural network was used only for choosing the best training samples, and was 
not employed in the final analysis ({\rm i.e.}, all events were reanalyzed).  
Removing events that were very unlikely $t\bar{t}$ candidates (NN$_0 < $ 0.3)
improved the efficiency of the training and increased network sensitivity to 
background events that more closely mimic $t\bar{t}$ event characteristics, 
thereby improving signal-to-background discrimination in the final analysis.

Training of the two neural networks used in the final analysis proceeded as 
follows.  The first neural network (NN$_1$) was trained on the ten kinematic 
variables using the training sets, as described above.  The output of NN$_1$, 
and the remaining three variables were then used as inputs to the second neural
network (NN$_2$).  NN$_2$ was trained using tagged {\sc herwig} $t\bar{t}$ 
Monte Carlo events and ``fake'' tagged data, also described in Sec.~VII.D.

%
%

%
\subsection{Cross section using neural network fits}
\label{mainmethod}
\begin{figure}
\centerline{\psfig{figure=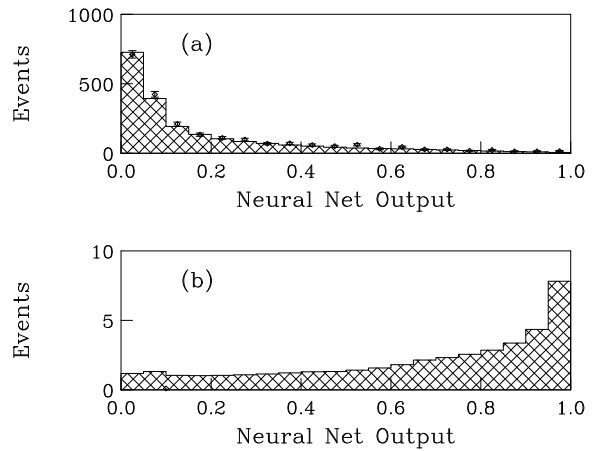,width=3in}}
\caption[Combined fits]
{\small The distributions in final neural network (NN$_2$) output for (a) 
data (diamonds) and expected background (histogram) and (b) {\sc herwig} 
$t\bar{t}$ signal for $m_t$=180~GeV/$c^2$.}
\label{comb_fit_lin}
\end{figure}

The $t\bar{t}$ cross section, integrated over all values of neural network 
output, is determined from the distributions in the output of the final neural 
network.  Any excess of the tagged data over the modeled background
distribution is attributed to $t\bar{t}$ production. This excess, integrated 
over all values of neural network output, is independent of the neural network,
and depends only on the accuracy of the modeling of the background by the tag 
rate function.  If the location of any excess appears in the region of 
$t\bar{t}$ signal (in neural network output) it would make these events likely 
$t\bar{t}$ candidates.  The final neural network (NN$_2$) distributions for 
the data and the expected background are shown in Fig.~\ref{comb_fit_lin}(a), 
and for {\sc herwig} $t\bar{t}$ events in Fig.~\ref{comb_fit_lin}(b). The 
normalization of the $t\bar{t}$ signal is described below. These distributions 
demonstrate a strong discrimination between signal and background.

\begin{figure}
\centerline{\psfig{figure=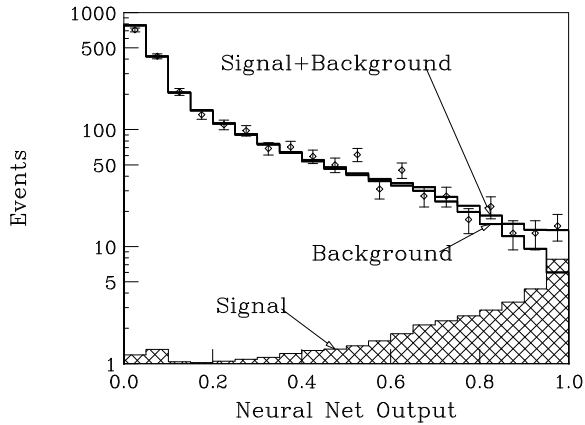,width=3in}}
\caption[Combined fits]
{\small The distribution in neural network (NN$_2$) output for data (diamonds)
and the fits for expected signal and background. The signal was modeled with 
{\sc herwig} for $m_t$=180~GeV/$c^2$. The errors shown are statistical.}
\label{comb_fit}
\end{figure}

We extract the cross section from a fit to the data of the sum of the neural
network output distributions expected for the $t\bar{t}$ signal and for QCD 
multijet background. Because the shapes of the $t\bar{t}$ and QCD network 
output distributions differ significantly, the relative amounts of each can be 
disentangled.  The generated {\sc herwig} $t\bar{t}$ events were arbitrarily
normalized assuming $\sigma_{t\bar{t}}$ = 6.4 pb at each top quark mass. This 
value needs to be factored out in normalizing Fig.~\ref{comb_fit_lin}(b).  
The data of Fig.~\ref{comb_fit_lin}(a) are fitted using $\chi^2$ minimization 
to the hypothesis:

\begin{equation}
N_{\rm expected} = {A_{\rm bkg}}~N^{i}_{\rm bkg} + 
               \frac{\sigma_{t\bar{t}}}{6.4~{\rm pb}}~N^{i}_{t\bar{t}},
\label{fitfun}
\end{equation}
\noindent
where $N^{i}_{\rm bkg}$ is the expected number of background events in the 
$i^{th}$ bin, and $N^{i}_{t\bar{t}}$ is the expected signal in this bin.
Because the full Monte Carlo sample, scaled to the total number of events 
(given by 6.4 pb multiplied by the integrated luminosity), is subjected to 
exactly the same trigger and selection criteria as the data, 
$N^{i}_{t\bar{t}}$ accounts for the luminosity, branching ratio (BR), 
and $t\bar{t}$ efficiency of our selection criteria. Both $A_{\rm bkg}$, 
the background normalization factor, and $\sigma_{t\bar{t}}$, are obtained 
from the fit, along with their respective statistical errors.  The results 
of this fit are shown in Fig.~\ref{comb_fit}.

By allowing the signal and background normalization factors to be determined 
from the fit, this method simultaneously provides the $t\bar{t}$ cross section 
and a more sensitive measurement of the background normalization. It 
efficiently exploits all information about the $t\bar{t}$ cross section and 
background normalization from the entire range of neural network output, 
without choosing any particular cutoff on neural network output. The 
distributions for signal, background and data are shown separately in 
Fig.~\ref{comb_fit}.  The error bars are the square root of the number of 
data events in each bin.

Events at the lowest values of neural network output ($<0.02$) have been 
removed, leaving 2207 events, or slightly more than half of the tagged data 
sample. The resulting fits may be checked by varying the region of NN$_2$ used.
(Fig.~\ref{comb_fit} uses events with NN$_2 >$ 0.02). Figure~27 shows results 
for $A_{\rm bkg}$ and $\sigma_{t\bar{t}}$ as a function of the lower limit in 
NN$_2$ employed in the fit. The results are seen to be quite stable to the 
change of this lower limit.  We note that the jets in events with NN$_2 <$ 0.02
tend to have low $E_T$, where the tagging rate may not be as well determined 
due to the low tagging probability. Because the background modeling may be 
less accurate in the very low NN$_2$ region, where the background so strongly 
dominates the data distribution, we impose a cut of NN$_2 >$ 0.02 for our fits 
to $A_{\rm bkg}$ and $\sigma_{t\bar{t}}$. The stability of the results shown 
in Fig. 27 supports this choice.

A similar plot was produced and fitted for several top quark masses, and the
values of the cross section obtained using the output distribution for 
{\sc herwig} $t\bar{t}$ events generated at that mass. The results are shown 
in Table VI for several top quark masses. Interpolating to the value for the 
top quark mass as measured by D\O~\cite{snydermass} ($m_t$ = 172.1 $\pm$ 7.1 
GeV), we obtain $\sigma_{t\bar{t}}$ = 7.1 $\pm$ 2.8(stat) pb.

Fitting the data in Fig.~\ref{comb_fit} only to the background 
($\sigma_{t\bar{t}}$ forced to zero), changes the normalization to 1.09 $\pm$ 
0.03, and the total $\chi^2$ per degree-of-freedom to 23.1/18.  We note that 
the change in $\chi^2$ comes predominantly from the last three bins of neural 
network output (in Fig.~\ref{comb_fit}), and the probability for a change in 
$\chi^2$ of 6.2 (for $m_t$=180~GeV/$c^2$) for one additional degree-of-freedom 
is consistent with the significance of the extracted cross section, which is 
2.5 standard deviations from zero.

\begin{figure}
\centerline{\psfig{figure=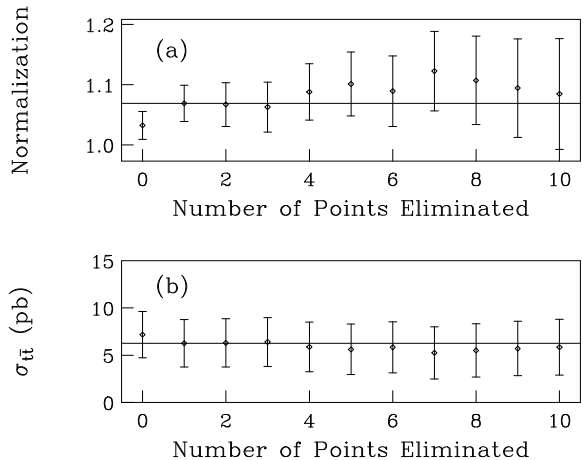,width=3in}}
\caption[Combined fit Errors]
{\small Results of combined fits (as in Fig.~\ref{comb_fit}) when data points 
are removed at small values of neural network output.  The refitted (a) 
background normalization and (b) $t\bar{t}$ cross section are plotted as a 
function of the number of points eliminated. Error bars are statistical, 
but are correlated through the error matrix.}
\label{comb_err}
\end{figure}

\begin{table}
 \begin{small}
 \caption{\small Results of the fits to neural network output.}
 \begin{center}
 \begin{tabular}{c|c|c|c}
 Top Quark               & $A_{\rm bkg}$   &  $\sigma_{t\bar{t}}$     
 & $\chi^2$ / d.o.f.  \\
 Mass (GeV/$c^2$)        &                 &  (pb)                    
 &                 \\
 \hline
 140   &  1.05 $\pm$ 0.03  &  18.4 $\pm$ 7.8  &  17.6 / 17 \\
 160   &  1.06 $\pm$ 0.03  &   9.3 $\pm$ 3.8  &  17.2 / 17 \\
 170   &  1.07 $\pm$ 0.02  &   7.2 $\pm$ 3.0  &  17.1 / 17 \\
 180   &  1.07 $\pm$ 0.03  &   6.3 $\pm$ 2.5  &  16.9 / 17 \\
 200   &  1.07 $\pm$ 0.03  &   5.1 $\pm$ 2.0  &  16.8 / 17 \\
 220   &  1.07 $\pm$ 0.03  &   4.4 $\pm$ 1.7  &  16.7 / 17 \\
 \end{tabular}
 \end{center}
 \end{small}
\end{table}
%
\subsection{Cross section using counting method}
\label{secondmethod}
The traditional method for extracting the $t\bar{t}$ cross section served as a 
useful check on the above procedure. We assumed an absolute normalization of 
the background as given by the tag rate function. Taking the excess in observed
events (seen in Fig.~\ref{comb_fit}) to be from $t\bar{t}$ production, we 
calculate the cross section for the process using the conventional relation:

\begin{equation}
\sigma_{t\bar{t}} =
        \frac{N_{\rm obs} - N_{\rm bkg}}{\epsilon \times BR \times
         {\cal{L}}  }
\label{convrel}
\end{equation}
\noindent
where {\it N}$_{\rm obs}$ is the number of observed events with neural network 
output greater than some threshold, {\it N}$_{\rm bkg}$ is the corresponding 
number of expected background events, $\epsilon \times BR$ is the branching 
ratio (BR) times the efficiency ($\epsilon$) of the criteria used for selecting 
$t\bar{t}$ events, and ${\cal{L}}$ is the total integrated luminosity 
(110.3~$\pm$ 5.8 pb$^{-1}$). We use {\sc herwig} as the model for calculating 
the value of $\epsilon \times BR$.

\begin{figure}
\centerline{\psfig{figure=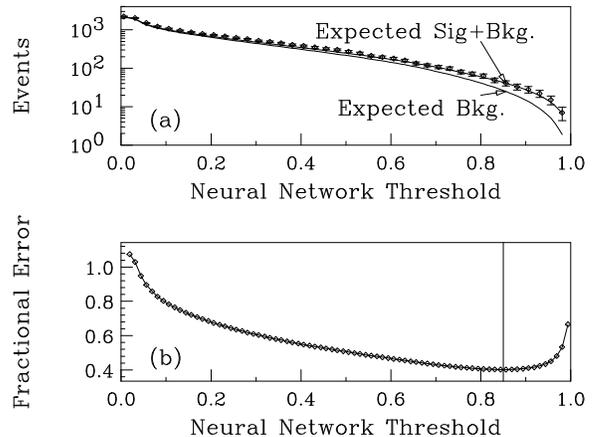,width=3in}}
\caption[PT fits]
{\small (a) The number of events (data) above any threshold on the neural 
network, and (b) the expected fractional error on the $t\bar{t}$ cross section
as a function of the threshold placed on the neural network output. The 
vertical line at \NNCUT~indicates the chosen threshold. The smooth curves in 
(a) represent the sum of the expected number of signal and background events
(assuming $m_t$=180 GeV/$c^2$ and $\sigma_{t\bar{t}}$=6.4 pb) and the expected 
number of background events only.}
\label{fin_nums}
\end{figure}

The number of events, as a function of the threshold placed on the output of 
the neural network, is shown in Fig.~\ref{fin_nums}(a).  The error bars are 
the square root of the number of events in each bin. The upper smooth curve in 
Fig.~\ref{fin_nums}(a) represents the sum of the expected signal and 
background, and the lower curve is just the expected background. The 
statistical error in the cross section depends upon where the threshold is 
placed. A plot of the relative statistical error versus the threshold on the 
output of the neural network is shown in Fig.~\ref{fin_nums}(b). The fractional
error ${\cal E}$ is approximated by:
\begin{equation}
{\cal E} = \frac{\sqrt{(N_{t\bar{t}}+N_{\rm bkg})}}{N_{t\bar{t}}},
\end{equation}
\noindent
where $N_{t\bar{t}}$ and $N_{\rm bkg}$ are the expected number of $t\bar{t}$ 
and background events above the neural network threshold. We wished to place 
the final threshold at or near the minimum error, and chose 0.85, as shown in 
Fig.~\ref{fin_nums}(b). The number of events above this threshold, the expected
background, and the expected signal are shown in Table VII.

\begin{table}
\caption[Fit results of muon tag rate.]
{\small Number of observed events, expected background, observed excess, and 
expected signal (assuming $m_t$=180 GeV/$c^2$ and $\sigma_{t\bar{t}}$=6.4 pb), 
for the threshold on the neural network output shown in Fig.~\ref{fin_nums}.}
\begin{small}
\begin{center}
\begin{tabular}{c|c|c|c}

Observed      &  Expected    & Observed    & Expected                 \\
Number        &  Background  & Excess      & {\sc herwig} $t\bar{t}$  \\
of Events     &  Events      & of Events   & Events \\
\hline
\NOBSERVED & \NBKG $\pm$ \DBKG & \EXCESS & 15.9 $\pm$ 2.6  \\
\end{tabular}
\end{center}
\end{small}
\end{table}

Using Eq.~\ref{convrel}, Table VIII lists the efficiency times branching ratios
for two input top quark mass values, and the extracted $t\bar{t}$ cross 
sections. We note that the method in Sec.~VII.F gave $t\bar{t}$ cross sections 
of 7.2 and 6.3 pb for $m_t$ of 170 and 180 GeV/$c^2$, respectively, in good 
agreement with the values in Table VIII. When interpolated to the measured top 
quark mass of 172.1 GeV/c$^2$, this determination yields a cross section of 
7.3 $\pm$ 3.0 $\pm$ 1.6 pb. The results from the fit to the neural network are 
slightly lower, as one would expect, since the background normalization was 
1.07 (instead of being fixed to 1 here). The changes in efficiencies as a 
function of top quark mass reflect the sensitivity of the selection criteria 
to the input mass $m_t$.  The statistical and systematic uncertainties in the 
cross sections are discussed in Sec.~VII.I.

\begin{table}
 \caption[Cross sections for $t\bar{t}$ ]
{\small Cross sections for $t\bar{t}$ production, using the counting method, 
obtained from the $b$-tagged events for $m_t$ = 170 and 180 GeV/$c^2$.}
 \begin{center}
  \begin{small}
  \begin{tabular}{c|c|c}
$m_t$       &    Signal Efficiency  & Cross Section \\
GeV/$c^2$   &    $\times$ BR        &   (pb)        \\
\hline
 170        &    0.019 $\pm$ 0.0032 &  7.5 $\pm$ 3.1 $\pm$ 1.6 \\
 180        &    0.022 $\pm$ 0.0037 &  6.5 $\pm$ 2.6 $\pm$ 1.4 \\
 \end{tabular}
 \end{small}
 \end{center}
\end{table}
%
\subsection{Double-tagged events}
The  requirement of a second  $b$-tagged jet  in the event  further reduces the
background, thereby  increasing the  signal-to-background ratio. Unfortunately,
the additional  requirement significantly reduces  the expected yield. However,
the search for these ``double-tagged''  events serves as a consistency check of
the single-tag  analysis, and also as  a test of the model  for the background.
The number of events that contain two  $b$-tagged jets is shown in Table IX for
various  NN$_2$  thresholds. The two  $b$-tags are  required to  originate from
separate jets; two  tags within the  same jet are counted  as a single tag. The
higher muon $p_T$ is used as the input to the neural network. The background is
again calculated based on Eq.~\ref{tagrate}, where 
$P_{\rm tag}(E_T,\eta,N_{\rm run})$, summed over all  jets, represents the 
expected number of tags  in the event.  The double-tag  probability is  
obtained via the Poisson  distribution, and is the  likelihood of observing  
at least two tagged jets, given the expected number. This  follows since the 
tag rate function is a rate per jet, and, within our model, the two tagged 
jets are uncorrelated.
 
We make the assumption that the fraction of double-tagged  events from
correlated sources, such as direct  heavy-quark pair production ($c \bar{c}$ or
$b \bar{b}$), remains unchanged over the  entire range of the neural network
output  variable.  This  assumption is  motivated by the fact that the energy
scales in such events are well above the energy thresholds for heavy-quark pair
production, and therefore the fraction of these events should be independent of
the neural network output. The good  agreement between the background model and
data in the single-tagged channel supports this assumption.

We determine the normalization of the  background by fitting the neural network
output distribution to  the expected background and  signal contributions as in
Sec.~VII.F.  The  32 events  were  binned in  neural  network  output,  and the
log-likelihood calculated. The minimum  in negative log-likelihood occurs for a
background normalization  factor of   0.97$^{+0.20}_{-0.18}$,  where the errors
correspond   to a  change  in    log-likelihood of  1/2.  In   determining this
normalization, the expected $t\bar{t}$ signal was not varied, but the result is
insensitive to  this value.  Allowing the data  to determine  the normalization
through this fit   accomodates the  possibility that the  tag rate function for
the second muon in the event is different from that for the first muon. The two
errors on the expected  background in Table IX  represent the uncertainties due
to the  tag  rate   function,  $t\bar{t}$   subtraction  and  $E_T$  scale (see
Sec.~VII.I)  and the normalization error, respectively. 

We note that the  fitted normalization  is consistent with  that for the single
tagged sample  indicating that the  second muon tag  probability is roughly the
same as for the first. The total  number of events for NN$_2 > 0.02$ is in good
agreement with the sum of expected  background plus the small contribution from
top.  The small  excess  persists  as the  NN$_2$  threshold  is  increased, in
agreement with  expectations. The  double tag analysis  supports our conclusion
that the singly-tagged sample is due to $t\bar{t}$ production.

\begin{table}
\caption[Fit results of muon 2tag rate.]
{\small Number of observed double-tagged events, expected background, observed 
excess, and expected signal (assuming $m_t$=180 GeV/$c^2$ and 
$\sigma_{t\bar{t}}$=6.4 pb), versus the threshold on the neural network output.
The first error in the expected background is due to the errors in the tag rate
function, $t\bar{t}$ correction, and the $E_T$ scale uncertainties. The second 
error is due to the uncertainty in the fitted background normalization factor, 
and is assumed to be fully correlated at different NN$_2$ values.}
\begin{small}
\begin{center}
\begin{tabular}{l|r|r|r|r}
NN$_2$        &  Observed   &  Expected    & Observed  & Expected            \\
Threshold     &  Number     &  Background  & Excess    
& {\sc herwig} $t\bar{t}$  \\
              &  of Events  &  Events      & of Events & Events \\
\hline
0.02           & 32         &  28.7 $\pm$ 5.5 $\pm$ 5.7   & 3.3      &  2.7 \\
0.1            & 22         &  16.6 $\pm$ 3.2 $\pm$ 3.3   & 5.4      &  2.7 \\
0.2            & 17         &  11.8 $\pm$ 2.3 $\pm$ 2.3   & 5.2      &  2.7 \\
0.4            & 12         &   6.8 $\pm$ 1.3 $\pm$ 1.4   & 5.2      &  2.5 \\
0.6            &  7         &   3.5 $\pm$ 0.7 $\pm$ 0.7   & 3.5      &  2.1 \\
0.8            &  3         &   1.1 $\pm$ 0.2 $\pm$ 0.2   & 1.9      &  1.4 \\
0.85           &  2         &   0.7 $\pm$ 0.1 $\pm$ 0.1   & 1.3      &  1.2 \\
0.9            &  1         &   0.4 $\pm$ 0.1 $\pm$ 0.1   & 0.6      &  1.0 \\
\end{tabular}
\end{center}
\end{small}
\end{table}
%
\subsection{Corrections and uncertainties}
\label{systematic}

In this subsection we discuss the major sources of systematic uncertainty that 
affect either the background estimate or signal efficiency.  The statistical
errors on the cross section and background normalization come directly from 
the fit (Eq.~\ref{fitfun}) shown in Fig.~\ref{comb_fit}.

\begin{itemize}

\item The statistical error in the calculation of the background is estimated 
by the number of untagged events falling in the signal region. This estimate 
of \NBKG~events, and an approximate mean tagging rate of 2\%, implies of the 
order of 1240 untagged events for the background, and a consequent 3\% 
statistical uncertainty in the background estimate. This contributes a 4\% 
uncertainty in the cross section based on the counting method in 
Eq.~\ref{convrel}.

\item The error in the normalization of the tagging rate was taken from the 
combined fits to the output of the neural networks using Eq.~\ref{fitfun}.  
This error is shown in Fig.~\ref{comb_err}(a), and was taken to be 5\%. It is 
used only in the calculation of the error on the background, as it is already 
included in the cross section. (The statistical error on the cross section was 
obtained from a simultaneous fit to the normalization of both background and 
signal, and accounts for the error on the background normalization.)

\item The uncertainty in the parameterization of the tagging rate results in a 
5\% uncertainty in the predicted number of background events. This was 
estimated by comparing the predicted number of tags for two functional forms 
(Eq.~\ref{cenmuonfit1} and Eq.~\ref{cenmuonfit2}) assumed for the tag rate.
Unlike the normalization of the tagging rate, this error accounts for possible 
changes in the shape of the background as a function of neural network output. 
 This results in a 7\% uncertainty in the $t\bar{t}$ cross section.

\item The presence of $t\bar{t}$ events in the data used for estimating 
background has been taken into account in all results presented thus far. The 
procedure used to estimate the correction to the background proceeds as 
follows. Calling $N_{t\bar{t}}^{\rm mistag}$ the number of untagged $t\bar{t}$ 
events wrongly assigned to the background estimate, we can estimate 
$N_{t\bar{t}}^{\rm mistag}$ as:
\begin{equation}
N_{t\bar{t}}^{\rm mistag} 
= \frac{0.8}{0.2} (N_{\rm obs} - N_{\rm bkg}) f_{\rm tag}
\end{equation}
where the $\frac{0.8}{0.2}$ corrects the $b$-tagged signal back to the untagged
signal (recall that $t\bar{t}$ events are tagged roughly 20\% of the time), 
$f_{\rm tag}$ is the average tag rate per event, and $N_{\rm obs}$ and 
$N_{\rm bkg}$ refer to events in the final tagged data sample. The corrected 
background estimation therefore becomes:
\begin{equation}
\label{nbkgcorr}
N_{\rm bkg}(corr) = N_{\rm bkg} - N_{t\bar{t}}^{\rm mistag}
\end{equation}
This correction is applied bin by bin in Fig.~\ref{comb_fit}, and is 
approximately 4\% in the signal region. We therefore assign a systematic 
uncertainty of 4\% to the background estimate and a corresponding 6\%
to the $t\bar{t}$ cross section.

\item Because untagged events, when multiplied by the tag rate function, model 
the tagged background, the $E_T$ scale of both sets must be the same.  
Any mismatch between these can produce subtle differences in the scales of 
the kinematic variables.  A useful measure of this scale is mean $H_T$.  
We observe that the difference in mean $H_T$ between our data and background 
model is 1.5 $\pm$ 1.4 GeV (see Fig.~\ref{muonpred}(a)), which is consistent
with no mismatch.  We take 1.4 GeV to be the uncertainty in the energy scale 
of the background model. This 1.4 GeV is added to one of the jets (we 
arbitrarily choose the jet with highest $E_T$), event-by-event, in the
background calculation and the analysis is redone.  The resultant change in 
the background is 4.2\%, and 9.1\% change in the cross section.

\item The statistical error in the $t\bar{t}$ efficiency is 3.2\%.

\item Any difference in the turn-on of the trigger efficiency for data and for 
$t\bar{t}$ Monte Carlo events can affect the signal efficiency. The difference 
can originate, for example, from the modeling of electronic noise or from the 
simulation of the underlying event. Furthermore, this efficiency can depend 
upon the mass of the top quark. From our trigger simulations, we estimate $<$ 
5\% uncertainty in signal efficiency from such 
sources \cite{Cathythesis,Eunilthesis}.

\item The uncertainty in the integrated luminosity was taken to be 
5.3\%~\cite{Bantly}.  This arises mainly from the uncertainty in the absolute 
luminosity, and affects all runs systematically.

\item Any difference in the relative energy scale between data and Monte Carlo 
affects the efficiency for signal. This uncertainty was determined using the 
MPF method\cite{mpf}, as described in Sec.~IV.C. Varying the energy scale in 
the $t\bar{t}$ Monte Carlo by $\pm$ (4\% + 1 GeV)~\cite{meenajim} changes the 
efficiency for signal by $\pm$ 5.7\%.

\item The $t\bar{t}$ tag rate is based on the $t\bar{t}$ Monte Carlo, but 
assumes that the performance of all detector components was stable during the 
run.  The Monte Carlo acceptance was reduced by 7.0\% to correct mainly for 
muon detection inefficiencies that were not modeled in our simulation. We 
estimate a 7.0\% uncertainty in the $t\bar{t}$ efficiency from any such 
changes in the muon tag rate.

\item Uncertainty in the model for $t\bar{t}$ production is estimated by 
comparing $t\bar{t}$ predictions from {\sc isajet} and {\sc herwig} generators.
Figure~\ref{effsys} shows the fractional differences in efficiencies 
(($\epsilon_{\sc isajet}$ - $\epsilon_{\sc herwig}$)/$\epsilon_{\sc herwig}$)
for different thresholds on $H_T$, $H_T^{3j}$, Aplanarity and \C~(again, for 
$m_t$ = 180 GeV/$c^2$). Although the two generators differ significantly in the
tails of these distributions, on average they are in reasonable agreement. The 
systematic error was estimated by repeating the analysis using events generated
 with {\sc isajet}. In order to remove the effects of the Fisher discriminant 
(${\cal F}$), which is not well modeled in {\sc isajet}, ${\cal F}$ values 
were randomly chosen based on a parameterization of the {\sc herwig} $t\bar{t}$ 
${\cal F}$ distribution. To further remove the dependence on the tag rate, 
randomly generated values of muon $p_T$ were taken. The expected distributions 
for the two generators, normalized as before, are shown in Fig.~\ref{isa_hw}. 
Identical thresholds were placed on the neural network output. The cross 
section changed by 6.2\%, which we take as the uncertainty in the overall 
signal efficiency due to $t\bar{t}$ model dependence.

\begin{figure}
\centerline{\psfig{figure=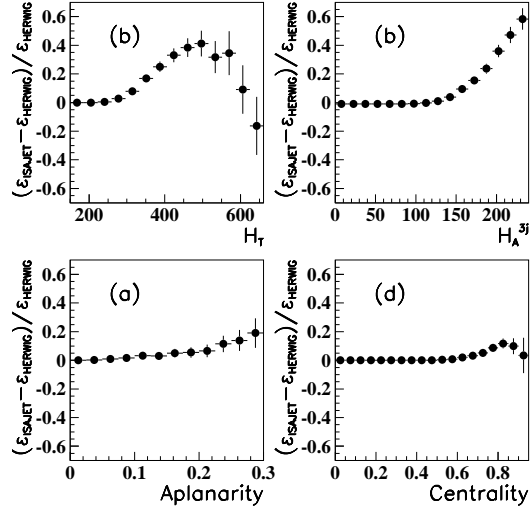,width=3in}}
\caption{\small Fractional differences in efficiencies between
{\sc isajet} and {\sc herwig}
($\epsilon_{\sc isajet}$ - $\epsilon_{\sc herwig}$)/$\epsilon_{\sc herwig}$
for $m_t$ = 180 GeV/$c^2$
(a) as a function of threshold on $H_T$, 
(b) as a function of threshold on $H_T^{3j}$,
(c) as a function of threshold on Aplanarity, and
(d) as a function of threshold on \C.}
\label{effsys}
\end{figure}

\begin{figure}
\centerline{\psfig{figure=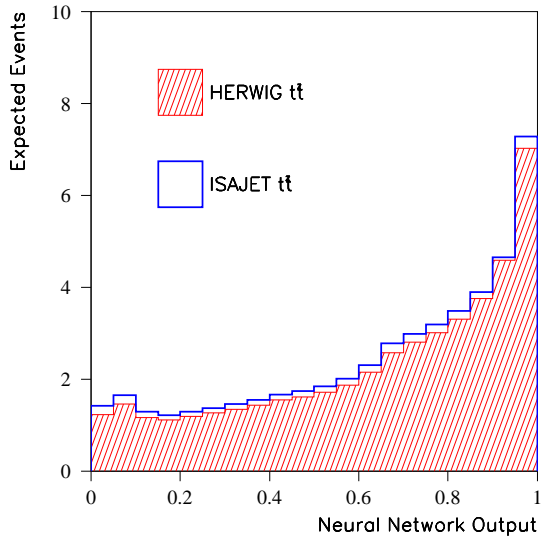,width=3in}}
\caption{\small Expected distributions in final neural network output (NN$_2$) 
for {\sc herwig} $t\bar{t}$ signal and {\sc isajet} $t\bar{t}$ signal for 
$m_t$=180~GeV/$c^2$.}
\label{isa_hw}
\end{figure}

\item The 6\% uncertainty in the $b\rightarrow\mu$ branching fraction~\cite{PDG}
corresponds to an average over the produced $B$-mesons. This 6\% enters directly
into the acceptance error in the Monte Carlo.

\item The $p_T$ of the tagged muon enters as an input to the neural network.
The mean $p_T$ in {\sc herwig} $t\bar{t}$ events was 14.7 GeV/$c$, while
in {\sc isajet} it was 15.9 GeV/$c$, an 8\% difference. Rescaling the muon 
$p_T$ in {\sc herwig} by 8\% changes the cross section by 7.0\%, which is 
taken as a systematic error.

\item The uncertainty resulting from the modeling of the Fisher discriminant 
for the jet widths, ${\cal F}$, was estimated by comparing data to our 
{\sc herwig} QCD Monte Carlo.  The mean value of ${\cal F}$ in data was 0.0470 
$\pm$ 0.0002 and in {\sc herwig} QCD it was 0.0488 $\pm$ 0.0019. The difference 
of 0.0018 $\pm$ 0.0019 indicates that our modeling is reasonable. The 
uncertainty on this result, 0.0019, was systematically added to the value 
of ${\cal F}$, event-by-event, in the {\sc herwig} $t\bar{t}$ generator, and 
the cross section recalculated. The observed change in the cross section of 
2.0\% is used as the systematic error from this variable.

\begin{table}
\caption[Summary of statistical and systematic uncertainties.]
{\small Summary of statistical and systematic uncertainties for the background 
estimate.}
\begin{center}
\begin{small}
\begin{tabular}{l|r}
Background Source & Size of Uncertainty   \\
\hline
Statistical Error                            &  3 \% \\
Normalization of the Muon Tag Rate           &  5 \% \\
Functional Form of the Muon Tag Rate         &  5 \% \\
Background Correction for $t\bar{t}$ Signal  & 4 \% \\
Background $E_T$ Scale                       & 4 \%   \\
\end{tabular}
\end{small}
\end{center}
\label{systematictable1}
\end{table}

\begin{table}
\caption[Summary of statistical and systematic uncertainties.]
{\small Summary of statistical and systematic uncertainties for the cross 
section.}
\begin{center}
\begin{small}
\begin{tabular}{l|r}
Background Source & Size of Uncertainty   \\
\hline
Statistical error                            &  4 \% \\
Functional Form of the Muon Tag Rate         &  7 \% \\
Background Correction for $t\bar{t}$ Signal  & 6 \% \\
Background $E_T$ scale                       & 9 \%   \\
\hline
Signal Source & Size of Uncertainty   \\
\hline
Statistical Error          & 3 \%  \\
Trigger Turn-on            & 5 \%  \\
Luminosity Error           & 5 \%  \\
Jet Energy Scale           & 6 \%  \\
$t\bar{t}$ Tag Rate        & 7 \%  \\
Model Dependence           & 6 \%  \\
$b\rightarrow\mu$ Branching Fraction     & 6 \%  \\
$p_T^{\mu}$ Dependence     & 7 \%  \\
${\cal{F}}$ Dependence     & 2 \%  \\
\end{tabular}
\end{small}
\end{center}
\label{systematictable2}
\end{table}

\end{itemize}

The sizes of the above effects are summarized in Table X for the uncertainties 
in the background, and in Table XI for the cross section. Adding both 
statistical and systematic errors in quadrature, we estimate the background 
as \NBKG~$\pm$ \DBKG~events (see Table VII). Similarly, the uncertainty in 
the efficiency of the \ttbar~signal is calculated from the errors in Table XI.

\subsection{Measured cross section}

\begin{figure}
\centerline{\psfig{figure=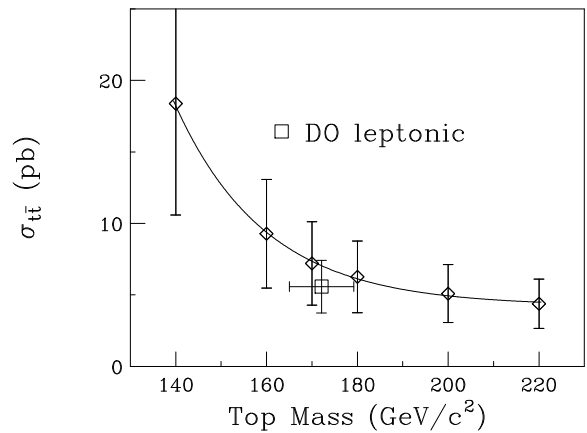,width=3in}}
\caption[allfits]
{\small The $t\bar{t}$ cross section extracted through fitting the shapes of 
the distributions in neural network output to data, shown as a function of top 
quark mass. Error bars are statistical only. For reference, the D\O\ $t\bar{t}$
cross section and top quark mass from leptonic channels~\cite{meenajim} is 
shown in the figure (open square).}
\label{allfits}
\end{figure}

By fitting the shape of the output in the neural network distribution,
we obtain the $t\bar{t}$ production cross section as a function of the 
input mass of the top quark. The $t\bar{t}$ cross sections extracted for 
several values of the top quark mass, along with a function used to 
interpolate the $t\bar{t}$ cross section (drawn as a smooth curve), are 
shown in Fig.~\ref{allfits}. Interpolating both the cross section and the 
statistical error, we find $\sigma_{t\bar{t}}$ = \XSEC~$\pm$ \DXSEC~$\pm$ 
\DXSECSYS~pb for $m_t$=\DOMASS~GeV/$c^2$~\cite{snydermass}.

The all-jets cross section can be combined with previous D\O~measurements of 
the \ttbar~production cross section, as extracted from channels where one or 
both of the $W$ bosons decay leptonically \cite{meenajim}. This cross section, 
averaged over all leptonic channels, was 5.6~$\pm$ 1.4~(stat) $\pm$ 
1.2~(syst) pb at $m_t$=\DOMASS~GeV/$c^2$, and is shown superimposed on 
Fig.~\ref{allfits}. The statistical errors on the all-jets and leptonic 
cross section measurements are uncorrelated. The systematic uncertainties in 
the following categories were assumed to be correlated with a correlation 
coefficient of 1.0.

\begin{itemize}
\item Luminosity.
\item Jet energy scale.
\item Muon tagging efficiency.
\item Non-leptonic trigger efficiency.
\item Top quark generator.
\item $b\rightarrow\mu$ branching ratio and muon $p_T$ spectrum.
\item Background tag rate function.
\end{itemize}
\noindent
The combined result for the D\O~\ttbar~production cross section is 
5.9~$\pm$ 1.2~(stat) $\pm$ 1.1 (syst)~pb for $m_t$=\DOMASS~GeV/$c^2$.

%
\subsection{Significance of signal}
\label{crosssection}

In this section, we estimate the significance of the excess of $t\bar{t}$ 
signal relative to expected background. We define the probability ($P$) of 
seeing at least the number of observed events ($N_{\rm obs}$), when only 
background is expected.  The significance of a $t\bar{t}$ signal can be 
characterized by the likelihood of $P$ being due to a fluctuation. If the 
distribution for the expected number of background events, $\mu$, is assumed
to be a Gaussian with mean $b$, and has a systematic uncertainty $\sigma_b$, 
then $P$ can be calculated as:

\begin{eqnarray}
P &=& \sum_{n=N_{\rm obs}}^{\infty}  \int^{\infty}_{0} d\mu
\frac{e^{-\mu}\mu^n}{n!}
\frac{1}{\sqrt{2\pi}\sigma_b} e^{-(\mu-b)^2/2\sigma_b^2} \nonumber \\
&=& 1 - \sum_{n=0}^{N_{\rm obs}-1} \int^{\infty}_{0} d\mu 
\frac{e^{-\mu}\mu^n}{n!}
\frac{1}{\sqrt{2\pi}\sigma_b} e^{-(\mu-b)^2/2\sigma_b^2}
\end{eqnarray}

\begin{figure}
\centerline{\psfig{figure=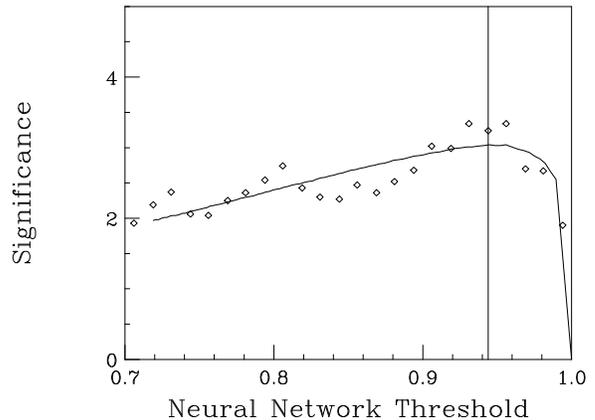,width=3in}}
\caption[significance]
{\small The expected (line) and observed (diamonds) values of significance of 
$t\bar{t}$ signal, plotted in terms of Gaussian equivalent standard deviations.
The vertical line corresponds to the cutoff that is expected to yield the 
greatest significance.}
\label{significance}
\end{figure}

The optimal choice of selection criteria can be found by minimizing the 
expected value of $P$ and, thereby, maximizing the significance of the excess, 
assuming that $N_{\rm obs}$ is composed of $t\bar{t}$ signal and background.  
Both the expected value and measured value of the significance are shown, along
with the cutoff for greatest significance, in Fig.~\ref{significance}. The 
result of the calculation, optimized for significance, with 18 observed events 
and an expected background of 6.9~$\pm$ 0.9, is $P$ = \psig, corresponding to 
a \sign~standard deviation effect. This is sufficient to establish the 
existence of a $t\bar{t}$ signal in multijet final states.

We consequently observe an excess in the multijet final states which we 
attribute to $t\bar{t}$ production. The cross section measured is consistent 
with previous measurements in other modes of $t\bar{t}$ decay\cite{meenajim}.
%
\section{Summary}
\label{concl}

We have performed a measurement of the \ttbar~production cross section in 
multijet final states. As described above, we observe an excess of events that 
can be attributed to $t\bar{t}$ production. The level of significance of the 
signal, as calculated from a possible upward fluctuation of the background to 
produce the observed excess, is sufficiently high to establish independently 
the existence of $t\bar{t}$ signal in the all-jets channel.

Using the D\O~measured value of \DOMASS~GeV/$c^2$ for the mass of the top 
quark, we obtain a cross section of \XSEC~$\pm$ \DXSEC~(stat) $\pm$ 
\DXSECSYS~(syst)~pb, which agrees with the D\O~cross section as measured in the
leptonic channels. Combining this result with previous D\O~measurements of the
\ttbar~production cross section gives 5.9~$\pm$ 1.2~(stat) $\pm$ 1.1 (syst)~pb.
%
\section*{Acknowledgement}
\label{ack}
We thank the staffs at Fermilab and collaborating institutions for their
contributions to this work, and acknowledge support from the 
Department of Energy and National Science Foundation (U.S.A.),  
Commissariat  \` a L'Energie Atomique (France), 
Ministry for Science and Technology and Ministry for Atomic 
   Energy (Russia),
CAPES and CNPq (Brazil),
Departments of Atomic Energy and Science and Education (India),
Colciencias (Colombia),
CONACyT (Mexico),
Ministry of Education and KOSEF (Korea),
and CONICET and UBACyT (Argentina).
%
%
%

%
\end{document}